\documentclass[aps,prd,preprint,showpacs,nofootinbib,eqsecnum,amsmath,amssymb]{revtex4-1}
\usepackage{epsf}
\usepackage{color}
\usepackage{dcolumn}
\usepackage{bm}
\everymath{\displaystyle}

\begin{document} 

\title{General treatment of quantum and classical spinning particles in external fields}

\author{\firstname{Yuri N.}~\surname{Obukhov}}
\email{obukhov@ibrae.ac.ru}
\affiliation{Theoretical Physics Laboratory, Nuclear Safety Institute,
Russian Academy of Sciences, B. Tulskaya 52, 115191 Moscow, Russia}

\author{\firstname{Alexander J.}~\surname{Silenko}}
\email{alsilenko@mail.ru} \affiliation{Research Institute for
Nuclear Problems, Belarusian State University, Minsk 220030, Belarus,\\
Bogoliubov Laboratory of Theoretical Physics, Joint Institute for Nuclear Research,
Dubna 141980, Russia}

\author{\firstname{Oleg V.}~\surname{Teryaev}}
\email{teryaev@theor.jinr.ru} \affiliation{Bogoliubov Laboratory
of Theoretical Physics, Joint Institute for Nuclear Research,
Dubna 141980, Russia,\\
National Research Nuclear University ``MEPhI'' (Moscow Engineering
Physics Institute), Kashirskoe Shosse 31, 115409 Moscow, Russia}


\begin {abstract}
We develop the general theory of spinning particles with electric and magnetic dipole moments moving in arbitrary electromagnetic, inertial and gravitational fields. Both the quantum-mechanical and classical dynamics is investigated. We start from the covariant Dirac equation extended to a spin-${\frac 12}$  fermion with anomalous magnetic and electric dipole moments and then perform the relativistic Foldy-Wouthuysen transformation. This transformation allows us to obtain the quantum-mechanical equations of motion for the physical operators in the Schr\"odinger form and to establish the classical limit of relativistic quantum mechanics. The results obtained are then compared to the general classical description of the spinning particle interacting with electromagnetic, inertial and gravitational fields.  The complete agreement between the quantum mechanics and the classical theory is proven in the general case. As an application of the results obtained, we consider the dynamics of a spinning particle in a gravitational wave and analyze the prospects of using the magnetic resonance setup to find possible manifestations of the gravitational wave on spin.
\end{abstract}

 \maketitle

\section{Introduction}

The motion of structureless particles \cite{Thir,Grei} under the action of inertial or gravitational forces and dynamics of charged point particles in the electromagnetic field \cite{Rohr,Kos} is well understood. The case of particles with microstructure (internal degrees of freedom, for example, spin) is more nontrivial. After Ulhenbeck and Goudsmit \cite{uhlen1,uhlen2} introduced the concept of spin to explain atomic spectra, Thomas \cite{Thomas1,Thomas2} and Frenkel \cite{frenkel1,frenkel2} promptly came up with the first models of particles with spin and magnetic moment (for a thorough historic overview, see \cite{Com}), and within the next year, Dirac \cite{Dirac1,Dirac2,Thal} formulated the relativistic quantum theory of a particle with spin ${\frac 12}$. Subsequently, considerable attention was paid to the study of the dynamics of charged spinning particles in the electromagnetic field, see \cite{bmt,Dixon,Suttorp}, to mention but a few notable references. One of the central issues was a comparison of the dynamics of spin in nonrelativistic classical mechanics, relativistic classical mechanics, nonrelativistic quantum mechanics, and relativistic quantum theory. An informative review of the results obtained can be found in the book of Corben \cite{corben}. 

Along with this, the motion of spinning particles in the gravitational (and/or inertial) field was extensively investigated. The fundamental issues of the spin-gravity coupling were studied and the corresponding methods were developed \cite{Tetrode,Weyl1,Weyl2,fock1,fock2,VDW,IVDW,Mathisson,Papapetrou,Bade,Riesz,Brill,peres0,Tiomno1,Schm}, as well as the numerous physical problems of spin dynamics were analysed \cite{fish,Huang,Ryder,barker,chicone,dinesh,kiefer} in the literature. The recent review \cite{NiRev} summarises the results obtained in this area and contains an exhaustive list of the relevant references. 

Much less attention was paid to the investigation of the spin dynamics under the joint action of the gravitational, inertial and electromagnetic field. Such a situation is of interest, on the one hand, as a fundamental problem of mathematical physics and, on the other hand, it has various important applications ranging form the astrophysical conditions (physical processes near the massive astrophysical objects like neutron stars or black holes) to the high-energy experimental setup on the Earth. In particular, our recent study \cite{OSTalkPRD} of the influence of terrestrial gravity and rotation in the precision experiments in storage rings has shown that the corresponding effects are not negligible, and they are manifest in perturbations of particle motion and in additional precession of spin. One should take them into account in the actual and planned $g$-2 and EDM experiments. 

In the present paper we study the most general case of the external gravitational and electromagnetic fields acting on a particle with microstructure (spin and dipole moments). A nontrivial feature of this system is in the absence of a direct superposition. The motion of a spinning particle only in the electromagnetic field or only in the gravitational field was investigated in the previous literature. However, when both external fields are present, their influence on spin and trajectory is not just a sum of two separate effects. This is explained by the well-known fact that, whereas electromagnetism couples only to electric charges and currents, gravity is universal and it couples to all types of matter, including the electromagnetic field. As a result, spinning particles feel the action of gravity both directly and via the electromagnetic field which gets modified in the curved spacetime. 

Our investigation is focused on the comparison of the classical and quantum spin dynamics, thereby generalizing Corben's analysis \cite{corben} from the purely electrodynamical setup to the case when a spinning particle moves under the combined action of arbitrary gravitational and electromagnetic fields. We continue here the development of the method of the Foldy-Wouthuysen (FW) for the quantum and semiclassical Dirac fermions which was started earlier for the spin dynamics in the special cases of weak and stationary field \cite{PRD,PRD2,Warszawa,OST} of a massive compact object, and further extended to the strong stationary \cite{OSTRONG} and arbitrary (strong and nonstatic) gravitational field \cite{ostgrav} in the Riemannian framework of Einstein's general relativity. Allowing for the nonminimal (Pauli-type dipole) coupling and for the possible deviations of the spacetime structure from the Riemannian geometry, we used our formalism to find new bounds on the spacetime torsion \cite{ostor,ostor2}. 

In this paper, we for the first time establish the complete consistency of the quantum, quasiclassical and classical spin dynamics for the Dirac fermion particles with dipole moments moving in arbitrary external gravitational plus electromagnetic field. This is our central result. Among numerous applications, we choose to briefly consider the motion of a spinning particle in the field of a gravitational wave and the magnetic field. 

The paper is organized as follows. In Sec. \ref{Prelim} we collect an introductory material, in particular, we describe the most general spacetime geometry and specify the corresponding coframe. The Hermitian Dirac Hamiltonian for the electrically charged fermion particle with dipole moments is derived in an arbitrary curved spacetime. Coordinate-free formulation Maxwell's electrodynamics on Riemannian manifolds is overviewed, and electric and magnetic fields are written down with respect to different coordinate (holonomic) and local Lorentz (anholonomic) frames, together with the Maxwell-Lorentz spacetime relation. After these preliminaries, in Sec. \ref{Hamiltonian} we present the basics of the FW transformation technique which is then applied to derive the exact FW Hamiltonian of the particle. The operator equations of motion are obtained, with the special attention to the dynamics (precession) of spin under the joint action of arbitrary gravitational and electromagnetic fields. The classical theory of spinning particle in external fields of any physical nature is overviewed in Sec. \ref{CS}, which is then subsequently specified in Sec. \ref{CSQ} to the case when the external fields encompass gravity and electromagnetism. The quantum and semiclassical spin dynamics is compared with the classical motion of a relativistic particle with dipole moments and the complete agreement of quantum and classical results is demonstrated. In order to illustrate how the formalism works, in Sec. \ref{Example} we consider the dynamics of a spinning particle in a general noninertial frame. Sec. \ref{SG} discusses a particle in the spacetime of a gravitational wave and magnetic field. Finally, the results obtained are summarized in Sec. \ref{Conclusion}. 

We use the same main conventions and notations as in Refs. \cite{OST,OSTRONG,ostgrav,MAG}. For the sake of completeness, let us remind that the world indices are labeled by Latin letters $i,j,k,\ldots = 0,1,2,3$ (for example, the local spacetime coordinates $x^i$ and the holonomic coframe $dx^i$), whereas we reserve Greek letters from the beginning of the alphabet for tetrad indices, $\alpha,\beta,\ldots = 0,1,2,3$ (e.g., the anholonomic coframe $\vartheta^\alpha$). Furthermore, spatial indices are denoted by Latin letters from the beginning of the alphabet, $a,b,c,\ldots = 1,2,3$. Note that in order to distinguish separate tetrad indices we put hats over them. We though omit hats for objects defined only in coframes. We use the standard mathematical symbols $\wedge$ and $^\ast$ for the exterior product and the Hodge duality operator, respectively. The metric of the Minkowski spacetime reads $g_{\alpha\beta} = {\rm diag}(c^2, -1, -1, -1)$, and the totally antisymmetric Levi-Civita tensor $\eta_{\alpha\beta\mu\nu}$ has the only nontrivial component $\eta_{\hat{0}\hat{1}\hat{2}\hat{3}} = c$, so that $\eta_{\hat{0}abc} = c\epsilon_{abc}$ with the three-dimensional Levi-Civita tensor $\epsilon_{abc}$. The spatial components ($a,b,\dots = 1,2,3$) of the tensor objects are raised and lowered with the help of the Euclidean 3-dimensional metric $\delta_{ab}$ (in some cases we write this out explicitly to avoid any misunderstanding). In the relativistic spinor theory, the four Dirac matrices $\gamma^\alpha$, $\alpha = 0,1,2,3$, satisfy the standard anticommutation condition $\gamma^\alpha\gamma^\beta + \gamma^\beta\gamma^\alpha = 2g^{\alpha\beta}$. As usual, $\sigma^{\alpha\beta} = {\frac i2}\left(\gamma^\alpha \gamma^\beta - \gamma^\beta\gamma^\alpha \right)$ are the generators of the local Lorentz transformations of the spinor field. For the Dirac matrices as well as for the gauge-theoretic notions and objects (including electrodynamics) we use the conventions of Bogoliubov-Shirkov \cite{BS}.

\section{Preliminaries}\label{Prelim}

In this section we present some basic facts about the geometry of the curved spacetime and the relativistic Dirac and Maxwell theories on Riemannian manifolds. 

Let $x^i = (t,x^a)$ be the local coordinates on the four-dimensional curved manifold. The spacetime interval
\begin{equation}
ds^2 = g_{ij}dx^i dx^j = g_{\alpha\beta}\vartheta^\alpha\vartheta^\beta\label{ds}
\end{equation}
can be equivalently written either in terms of the holonomic coframe $dx^i$ or in terms of the anholonomic (tetrad) one $\vartheta^\alpha = e^\alpha_idx^i$. We describe the components of the latter in the Schwinger gauge $e_a^{\,\widehat{0}} =0$ (also $e_{\widehat{a}}^{\,0} =0), ~a=1,2,3$, as follows:
\begin{equation}\label{coframe}
e_i^{\,\widehat{0}} = V\,\delta^{\,0}_i,\qquad e_i^{\widehat{a}} =
W^{\widehat a}{}_b\left(\delta^b_i - cK^b\,\delta^{\,0}_i\right),\qquad a=1,2,3.
\end{equation}
Here we assume that the functions $V = V(x^i)$ and $K^a = K^a(x^i)$, as well as the components of the $3\times 3$ matrix $W^{\widehat a}{}_b = W^{\widehat a}{}_b(x^i)$ may depend arbitrarily on the local coordinates $t,x^a$.

The coframe (\ref{coframe}) gives rise to a general form of the spacetime line element (\ref{ds})
\begin{equation}\label{LT}
ds^2 = V^2c^2dt^2 - \delta_{\widehat{a}\widehat{b}}W^{\widehat a}{}_c W^{\widehat b}{}_d
\,(dx^c - K^ccdt)\,(dx^d - K^dcdt).
\end{equation}
This is the well-known Arnowitt-Deser-Misner (ADM) parametrization of the metric \cite{ADM} which we previously used in Ref. \cite{OSTRONG}. The components of the spacetime metric $g_{ij}$ and of its inverse $g^{ij}$ read explicitly
\begin{eqnarray}
g_{00} = c^2(V^2 - \underline{g}{}_{ab}K^aK^b),\qquad g_{0a} = c\underline{g}{}_{ab}K^b,\qquad
g_{ab} = -\,\underline{g}{}_{ab},\label{gij}\\
g^{00} = {\frac 1{c^2V^2}},\qquad g^{0a} = {\frac {K^a}{cV^2}},\qquad
g^{ab} = -\,\underline{g}{}^{ab} + {\frac 1{V^2}}\,K^aK^b.\label{ijg}
\end{eqnarray}
Here the spatial 3-dimensional metric is given by $\underline{g}{}_{ab} = \delta_{\widehat{c}\widehat{d}}W^{\widehat c}{}_a W^{\widehat d}{}_b$, and $\underline{g}{}^{ab} = \delta^{\widehat{c}\widehat{d}}W^a{}_{\widehat c} W^b{}_{\widehat d}$. The $3\times 3$ matrix ${W}^b{}_{\widehat a}$ is inverse to ${W}^{\widehat a}{}_b$. The off-diagonal metric components $g_{0a}$ and $g^{0a}$ are related to the effects of rotation.

For the ADM parametrization (\ref{LT}) of the general spacetime metric with the tetrad (\ref{coframe}), the components of the local Lorentz connection read explicitly
\begin{eqnarray}
\Gamma_{i\,\widehat{a}\widehat{0}} &=& {\frac {c^2}V}\,W^b{}_{\widehat{a}}
\,\partial_bV\,e_i{}^{\widehat{0}} - {\frac cV}\,{\cal Q}_{(\widehat{a}
\widehat{b})}\,e_i{}^{\widehat{b}},\label{connection1}\\
\Gamma_{i\,\widehat{a}\widehat{b}} &=& {\frac cV}\,{\cal Q}_{[\widehat{a}
\widehat{b}]}\,e_i{}^{\widehat{0}} + \left({\cal C}_{\widehat{a}\widehat{b}
\widehat{c}} + {\cal C}_{\widehat{a}\widehat{c}\widehat{b}} + {\cal C}_{\widehat{c}
\widehat{b}\widehat{a}}\right) e_i{}^{\widehat{c}},\label{connection2}
\end{eqnarray}
where we introduced (denoting the partial derivative with respect
to the coordinate time $t$ by the dot $\dot{\,} = \partial_t$\,)
\begin{eqnarray}
{\cal Q}_{\widehat{a}\widehat{b}} &=& g_{\widehat{a}\widehat{c}}W^d{}_{\widehat{b}}
\left({\frac 1c}\dot{W}^{\widehat c}{}_d + K^e\partial_e{W}^{\widehat c}{}_d
+ {W}^{\widehat c}{}_e\partial_dK^e\right),\label{Qab}\\
{\cal C}_{\widehat{a}\widehat{b}}{}^{\widehat{c}} &=& W^d{}_{\widehat{a}}
W^e{}_{\widehat{b}}\,\partial_{[d}W^{\widehat{c}}{}_{e]},\qquad {\cal
C}_{\widehat{a} \widehat{b}\widehat{c}} = g_{\widehat{c}\widehat{d}}
\,{\cal C}_{\widehat{a}\widehat{b}}{}^{\widehat{d}}.\label{Cabc}
\end{eqnarray}
We can obviously identify ${\cal C}_{\widehat{a}\widehat{b}}{}^{\widehat{c}} =- {\cal C}_{\widehat{b}\widehat{a}}{}^{\widehat{c}}$ with the reduced anholonomity object for the spatial triad ${W}^{\widehat a}{}_b$.

In order to give the most general description of electromagnetic interactions of a Dirac particle, we allow for the nonminimal coupling with the possible anomalous dipole moments of the particle. Accordingly, the covariant Dirac equation for the spinor field $\Psi$ with the mass $m$, the anomalous magnetic moment (AMM) $\mu'$ and the electric dipole moment (EDM) $\delta'$ reads \cite{ostor}:
\begin{equation}
\left(i\hbar\gamma^\alpha D_\alpha - mc + {\frac{\mu'}{2c}}\sigma^{\alpha\beta}F_{\alpha\beta}
+ {\frac{\delta'}{2}}\sigma^{\alpha\beta}G_{\alpha\beta}\right)\Psi=0.\label{Diracgen}
\end{equation}
The spinor covariant derivative
\begin{equation}
D_\alpha = e_\alpha^i D_i,\qquad D_i = \partial _i - {\frac {ie}{\hbar}}
\,A_i + {\frac i4}\sigma^{\alpha\beta}\Gamma_{i\,\alpha\beta},\label{eqin2}
\end{equation}
describes the minimal interaction of a fermion particle with the external classical fields: the electromagnetic 4-potential $A_i$ (coupled to the electric charge $e$ of a fermion), and the gravitational field potentials $(e^\alpha_i, \Gamma_i{}^{\alpha\beta})$. The tetrad indices of the Dirac matrices reflect the definition of the three-component physical spin (pseudo)vector in the local Lorentz rest frame of a particle. In the limit of the flat Minkowski spacetime, Eq. (\ref{Diracgen}) reduces to the Dirac-Pauli equation for a particle with the AMM and EDM (see Refs. \cite{Com,RPJ}).

The tensors $F_{\alpha\beta}$ and $G_{\alpha\beta}$ in (\ref{Diracgen}) are defined as $F_{\alpha\beta}=e_\alpha^i e_\beta^j F_{ij}$ and $G_{\alpha\beta} =e_\alpha^i e_\beta^j G_{ij}$, where $F_{ij} = \partial_i A_j - \partial_j A_i$ is the electromagnetic field strength tensor and its Hodge dual is $G_{ij}={\frac 12}\eta_{ijkl}F^{kl}$.

We can recast the Dirac equation (\ref{Diracgen}) into the Schr\"odinger form, however, the corresponding ``naive'' Hamiltonian is non-Hermitian \cite{GorNezn,GorNeznnew,lec}. This problem is solved by rescaling of the spinor wave function $\psi = (\sqrt{-g}e^0_{\widehat 0})^{\frac 12}\Psi$, and the resulting Schr\"odinger equation $i\hbar\frac{\partial \psi} {\partial t}= {\cal H}\psi$ then contains the Hermitian (and self-adjoint) Hamiltonian
\begin{eqnarray}
{\cal H} &=& \beta mc^2V + e\Phi + {\frac c 2}\left(\pi_b\,{\cal F}^b{}_a \alpha^a
+ \alpha^a{\cal F}^b{}_a\pi_b\right) + {\frac c2}\left(\bm{K}\!\cdot\bm{\pi}
+ \bm{\pi}\!\cdot\!\bm{K}\right)\nonumber\\ \label{HamiltonDP}
&& +\,{\frac {\hbar c}4}\left(\bm{\Xi}\!\cdot\!\bm{\Sigma} - \Upsilon\gamma_5\right)
- \beta V\left(\bm{\Sigma}\cdot\bm{\mathcal M} + i\bm{\alpha}\cdot\bm{\mathcal P}\right).
\end{eqnarray}
Here, as usual,  $\alpha^a = \beta\gamma^a$ ($a,b,c,\dots = 1,2,3$) and the spin matrices
$\Sigma^1 = i\gamma^{\hat 2}\gamma^{\hat 3}, \Sigma^2 = i\gamma^{\hat 3}\gamma^{\hat 1},
\Sigma^3 = i\gamma^{\hat 1}\gamma^{\hat 2}$ and $\gamma_5=i\alpha^{\hat{1}}\alpha^{\hat{2}}
\alpha^{\hat{3}}$. Boldface notation is used for 3-vectors ${\bm K} = \{K^a\},\, {\bm\alpha}
= \{\alpha^a\}, \,{\bm\Sigma} = \{\Sigma^a\},\,{\bm\pi} = \{\pi_a\}$. The latter is the kinetic momentum operator, $\bm\pi=-i\hbar\bm{\nabla} - e\bm A$. The minimal coupling gives rise to the terms in (\ref{HamiltonDP}) with the objects 
\begin{eqnarray}\label{AB1}
{\cal F}^b{}_a &=& VW^b{}_{\widehat a},\\
\Upsilon &=& V\epsilon^{\widehat{a}\widehat{b}\widehat{c}}\Gamma_{\widehat{a}\widehat{b}\widehat{c}} =
- V\epsilon^{\widehat{a}\widehat{b}\widehat{c}}{\cal C}_{\widehat{a}\widehat{b}\widehat{c}},\label{AB2}\\
\Xi^a &=& {\frac Vc}\,\epsilon^{\widehat{a}\widehat{b}\widehat{c}}\Gamma_{\widehat{0}\widehat{b}\widehat{c}}
= \epsilon_{\widehat{a}\widehat{b}\widehat{c}}\,{\cal Q}^{\widehat{b}\widehat{c}},\label{AB3}
\end{eqnarray}
whereas the nonminimal coupling is encoded in
\begin{equation}\label{MaPa}
\bm{\mathcal M}^a = \mu'\bm{\mathfrak B}^a + \delta'\bm{\mathfrak E}^a,\qquad
\bm{\mathcal P}_a = c\delta'\bm{\mathfrak B}_a - \mu'\bm{\mathfrak E}_a/c.
\end{equation}

Now, let us recall the basics of the classical electrodynamics on curved manifolds. We should carefully distinguish the {\it anholonomic} components $\bm{\mathfrak{E}}, \bm{\mathfrak{B}}$ of the Maxwell tensor, $\bm{\mathfrak{E}}{}_a = \{ F_{\widehat{1}\widehat{0}}, F_{\widehat{2}\widehat{0}}, F_{\widehat{3}\widehat{0}} \}$ and $\bm{\mathfrak{B}}{}^a = \{ F_{\widehat{2}\widehat{3}}, F_{\widehat{3}\widehat{1}}, F_{\widehat{1}\widehat{2}} \}$, and the {\it holonomic} components $\bm{E}, \bm{B}$ of $F_{ij}$, which are $\bm{E}_a = \{ F_{10}, F_{20}, F_{30} \}$ and $\bm{B}^a = \{ F_{23}, F_{31}, F_{12} \}$. For the general metric (\ref{LT}) with the tetrad (\ref{coframe}), these fields are related via (denoting $w := \det W^{\widehat c}{}_d$)
\begin{eqnarray}\label{EE}
\bm{\mathfrak{E}}{}_a &=& {\frac 1V}\,{W}^b{}_{\widehat a}(\bm{E} + c\bm{K}\times\bm{B})_b,\\
\bm{\mathfrak{B}}{}^a &=& {\frac 1{w}}\,W^{\widehat a}{}_b\,\bm{B}^b, \label{BB}
\end{eqnarray}
Hereafter the vector product is defined by $\{\bm A\times\bm B\}_a=\epsilon_{abc}A^bB^c$ for any 3-vectors $A^b$ and $B^c$. The dynamics of the electromagnetic field is described in terms of the field strength 2-form $F = {\frac 12}F_{ij}dx^i\wedge dx^j$ and the electromagnetic excitation 2-form $H = {\frac 12}H_{ij}dx^i\wedge dx^j$. These fundamental variables satisfy the Maxwell equations. The latter are written in a generally covariant form which is valid for all coordinates and reference frames \cite{HehlObukhov}:
\begin{eqnarray}
dF &=& 0,\label{maxF0}\\ dH &=& J.\label{maxH0}
\end{eqnarray}
The current 3-form $J  = {\frac 16}J_{ijk}dx^i\wedge dx^j\wedge dx^k$ describes the distribution of the electric charges and currents which are the sources of the electromagnetic field. To make the theory predictive, the system (\ref{maxF0})-(\ref{maxH0}) should be supplemented by the constitutive relations between $F$ and $H$. In the Maxwell-Lorentz electrodynamics, the constitutive relation reads
\begin{equation}
H = \lambda_0\star\!F,\qquad \lambda_0 = \sqrt{\frac {\varepsilon_0}{\mu_0}}.\label{const}
\end{equation}
There $\varepsilon_0$ and $\mu_0$ are the electric and magnetic constants of the vacuum (not to confuse the latter with the magnetic dipole moment!), and the star $\star$ denotes the Hodge duality operator determined by the spacetime metric.

The equations (\ref{maxF0}), (\ref{maxH0}) and (\ref{const}) can be written in the equivalent vector form, see in Ref. \cite{HehlItinObukhov} for more details. Introducing the components of the magnetic and electric excitations, $\bm{H}_a = \{ H_{01}, H_{02}, H_{03} \}$ and $\bm{D}^a = \{ H_{23}, H_{31}, H_{12} \}$, and identifying the components of the source 3-form $J$ with the electric current density $\bm{J}^a = \{-\,J_{023}, -\,J_{031}, -\,J_{012} \}$ and the charge density $\rho = J_{123}$, we recast the Maxwell equations (\ref{maxF0})-(\ref{maxH0}) into \cite{HehlItinObukhov}
\begin{eqnarray}
\bm{\nabla}\times \bm{E} + \dot{\bm{B}} = 0,
\qquad \bm{\nabla}\cdot\bm{B} = 0,\label{maxFR}\\
\bm{\nabla}\times \bm{H} - \dot{\bm{D}} = \bm{J},
\qquad \bm{\nabla}\cdot\bm{D} = \rho.\label{maxHR}
\end{eqnarray}
The influence of the inertia and gravity is encoded in the Maxwell-Lorentz constitutive law (\ref{const}). The latter can be recast into the explicit constitutive relations between the components of electric and magnetic fields $\bm{E}, \bm{B}$ and the electric and magnetic excitations $\bm{D}, \bm{H}$:
\begin{eqnarray}
D^a &=& {\frac {\varepsilon_0w}{V}}\,\underline{g}{}^{ab}E_b
- \lambda_0{\frac {w}{V}}\,\underline{g}{}^{ad}\epsilon_{bcd}K^c\,B^b,\label{const1}\\
H_a &=& {\frac {1}{\mu_0wV}}\left\{(V^2 - K^2)\underline{g}{}_{ab} + K_aK_b\right\}B^b
-\,\lambda_0{\frac {w}{V}}\,\epsilon_{adc}K^c\,\underline{g}{}^{db}E_b.\label{const2}
\end{eqnarray}
Here $K_a = \underline{g}{}_{ab}K^b$, and $K^2 = K_aK^a = \underline{g}{}_{ab}K^aK^b$ (and recall that $w = \det W^{\widehat c}{}_d$).

\section{Foldy-Wouthuysen transformation for a Dirac particle}\label{Hamiltonian}

In order to reveal the physical contents of the Schr\"odinger equation, we need to go to the Foldy-Wouthuysen (FW) \cite{FW} representation. Earlier \cite{OST,OSTRONG,ostgrav} we considered the purely gravitational case by dropping the terms depending on the electromagnetic field. Here we turn to the general case and take into account both gravity and electromagnetism. We can construct the FW transformation \cite{FW} for the Dirac Hamiltonian (\ref{HamiltonDP}) with the help of the general method developed in Refs. \cite{PRA,PRAnonstat,PRA2015}. It is worthwhile to mention that there is a lot of different approaches to the FW transformation (see Refs. \cite{JMPcond,PRA2016,E,Expooperator} and references therein). The use of the method \cite{PRA,PRAnonstat,PRA2015} allows us to derive the FW Hamiltonian which is exact in all terms of the zero and the first orders in the Planck constant $\hbar$ and which also includes the second order terms in the Planck constant which describe contact interactions. 

All the resulting quantum-mechanical Hamiltonians and the equations of motion are Hermitian. To avoid quite cumbersome expressions, we will neglect noncommutativity of the coordinate and momentum operators in some formulas, since the apparent corrections for a non-Hermitian form of the corresponding terms are always negligible. In our calculations, we take into account only terms of the first order in $\bm{\Xi},\Upsilon,\bm{\mathcal P},\bm{\mathcal M}$ and neglect their higher powers and bilinear combinations. Omitting the technical details (see \cite{OST,OSTRONG,ostgrav,ostor,ostor2} for the description of computational methods) we find for the FW Hamiltonian:
\begin{equation}
{\cal H}_{FW}={\cal H}_{FW}^{(1)}+{\cal H}_{FW}^{(2)}+{\cal H}_{FW}^{(3)}+{\cal H}_{FW}^{(4)}.\label{eqFW}
\end{equation}
The four terms in this sum read, respectively,
\begin{eqnarray}
{\cal H}_{FW}^{(1)} &=& \beta\epsilon' + \frac{\hbar c^2}{16}\left\{
\frac{1}{\epsilon'},\left(2\epsilon^{cae}\Pi_e \{\pi_b,{\cal F}^d{}_c\partial_d
{\cal F}^b{}_a\}+\Pi^a\{\pi_b,{\cal F}^b{}_a\Upsilon\}\right)\right\}\nonumber\\
&& +\,\frac{\hbar mc^4}{4}\epsilon^{cae}\Pi_e\left\{\frac{1}{{\cal T}},\left\{
\pi_d,{\cal F}^d{}_c{\cal F}^b{}_a\partial_bV\right\}\right\},\label{eq7}\\
{\cal H}_{FW}^{(2)} &=& \frac c2\left(K^a \pi_a + \pi_a K^a\right)
+ {\frac {\hbar c}4}\,\Sigma_a\Xi^a\nonumber\\
&& +\,\frac{\hbar c^2}{16}\Biggl\{\frac{1}{{\cal T}},\biggl\{\Sigma_a \{\pi_e,
{\cal F}^e{}_b\},\Bigl\{\pi_f,\bigl[\epsilon^{abc}({\frac 1c} \dot{\cal F}^f{}_c 
- {\cal F}^d{}_c\partial_dK^f + K^d\partial_d{\cal F}^f{}_c)\nonumber\\
&& -\,{\frac 12}{\cal F}^f{}_d\left(\delta^{db}\Xi^a -
\delta^{da}\Xi^b\right) \bigr]\Bigr\}\biggr\}\Biggr\},\label{eq7K}\\
{\cal H}_{FW}^{(3)} &=& e\Phi - {\frac{e\hbar c^2}{4}}\left\{\frac{1}{\epsilon'},V^2\Pi^a
{\mathfrak B}_a\right\}\nonumber\\
&& +\,\frac{e\hbar c^2}{8}\left\{\frac{1}{{\cal T}}, \Bigl[\Sigma_a\epsilon^{abc}\left(
\{{\cal F}^d{}_b,\pi_d\}V^2{\mathfrak E}_c - V^2{\mathfrak E}_b\{{\cal F}^d{}_c,\pi_d\}\right) 
- 2\hbar{\cal F}^b{}_a\partial_b(V^2{\mathfrak E}^a)\Bigr]\right\},\label{eq73}\\
{\cal H}_{FW}^{(4)} &=& -\,{\frac c8}\biggl\{\frac{1}{\epsilon'},\Bigl[\Sigma_a
\epsilon^{abc}\bigl(\{{\cal F}^d{}_b,\pi_d\}V{\mathcal{P}}_c - V{\mathcal{P}}_b\{{\cal F}^d{}_c,
\pi_d\}\bigr) - 2\hbar{\cal F}^b{}_a\partial_b(V{\mathcal{P}}^a)\Bigr]\biggr\} 
- V\,\Pi^a{\mathcal{M}}_a\nonumber\\
&& +\,\frac{c^2}{4}\biggl\{\frac{1}{{\cal T}},\Bigl(\Pi^a\bigl\{\{{\cal F}^c{}_a{\cal F}^d{}_b
V{\mathcal{M}}^b,\pi_c\},\pi_d\bigr\}+\beta\hbar\left\{{\cal F}^b{}_a[\mathcal{J}^a
+ K^c\partial_c(V{\mathcal{P}}^a)],\pi_b\right\}\Bigr)\biggr\}.\label{eq74}
\end{eqnarray}
Here we introduced the operators
\begin{eqnarray}
\epsilon' = \sqrt{m^2c^4V^2+\frac{c^2}{4}\delta^{ac}\{\pi_b,{\cal F}^b{}_a\}
\{\pi_d,{\cal F}^d{}_c\}},\qquad {\cal T}=2{\epsilon'}^2 + \{\epsilon',mc^2V\},\nonumber\\
\mathcal{J}^a = \epsilon^{abc}{\cal F}^d{}_b \partial_d(V\mathcal{M}_c) 
+ {\frac {\partial\mathcal{P}^a}{c\partial t}}.\label{eqa}
\end{eqnarray} 
We do not include the term which also results from our derivations,
\begin{equation}
-\beta\frac{e\hbar c}{2}\biggl\{\frac{1}{{\cal T}},V^3\bm{\mathcal{P}}\cdot\bm{\mathfrak{E}}\biggr\},
\nonumber
\end{equation}
into the FW Hamiltonian because it does not describe any contact interaction. The FW Hamiltonian (\ref{eqFW}) is Hermitian and self-adjoint. The first two terms (\ref{eq7}) and (\ref{eq7K}) determine the dynamics of the Dirac fermion on the Riemannian spacetime manifold, whereas (\ref{eq73}) and (\ref{eq74}) give the general description of the contribution of the electromagnetic field to the FW Hamiltonian (accounting for the minimal and non-minimal interaction, respectively). In the absence of electromagnetic field, we recover the previous results \cite{Ob1,Ob2,PRD,PRD2,Warszawa,OST,OSTRONG,ostgrav}. As compared to Ref. \cite{ostgrav}, Eqs. (\ref{eq7}) and (\ref{eq7K}) differ by the replacement of $\bm p =-i\hbar\bm{\nabla}$ with $\bm\pi = - i\hbar\bm{\nabla} - e\bm A$. Equations (\ref{eq73}) and (\ref{eq74}) agree with the corresponding equations in quantum electrodynamics \cite{RPJ}. Remarkably, the Hamiltonian (\ref{eqFW})-(\ref{eq74}) contains only anholonomic fields $\bm{\mathfrak{E}}$ and $\bm{\mathfrak{B}}$ in the spin-dependent terms.

To analyse the dynamics of the spin, we need to evaluate the commutator of the polarization operator $\bm\Pi=\beta\bm\Sigma$ with the FW Hamiltonian (\ref{eqFW}). The derivation is straightforward and results in the dynamical equation that describes the precession of the spin in the exterior gravitational and electromagnetic fields (cf. Ref. \cite{ostgrav}):
\begin{equation}
\frac{d\bm \Pi}{dt}=\frac{i}{\hbar}[{\cal H}_{FW},\bm \Pi]=\bm\Omega_{(1)}
\times\bm \Sigma+\bm\Omega_{(2)}\times\bm\Pi.\label{spinmeq}
\end{equation}
The components of the 3-vector operators of the angular velocity $\bm\Omega_{(1)}$ and $\bm\Omega_{(2)}$ are as follows:
\begin{eqnarray}
\Omega^a_{(1)} &=& \frac{mc^4}{2}\left\{{\frac 1{\cal T}}, \{\pi_e,
\epsilon^{abc}{\cal F}^e{}_b{\cal F}^d{}_c\partial_d\,V\}\right\}\nonumber\\
&& +\,\frac{c^2}{8}\left\{\frac{1}{\epsilon'},\{\pi_e,(2\epsilon^{abc}
{\cal F}^d{}_b\partial_d{\cal F}^e{}_c + \delta^{ab}
{\cal F}^e{}_b\,\Upsilon)\} \right\}\nonumber\\
&& +\,\frac{ec^2}{4}\epsilon^{abc}\left\{\frac{1}{{\cal T}},\left(\{{\cal F}^d{}_b,\pi_d\}V^2
{\mathfrak E}_c - V^2{\mathfrak E}_b\{{\cal F}^d{}_c,\pi_d\}\right) \right\}\nonumber\\
&& -\,{\frac c{4\hbar}}\epsilon^{abc}\biggl\{\frac{1}{\epsilon'},\bigl(\{{\cal F}^d{}_b,\pi_d\}V{\mathcal{P}}_c - V{\mathcal{P}}_b\{{\cal F}^d{}_c,\pi_d\}\bigr)\biggr\},\label{eqol}
\end{eqnarray}
and
\begin{eqnarray}
\Omega^a_{(2)} &=& \frac{c^2}{8}\Biggl\{\frac{1}{{\cal T}},
\biggl\{ \{\pi_e,{\cal F}^e{}_b\},\Bigl\{\pi_f,\bigl[\epsilon^{abc}
({\frac 1c} \dot{\cal F}^f{}_c - {\cal F}^d{}_c\partial_dK^f
+ K^d\partial_d{\cal F}^f{}_c)\nonumber\\
&& -\,{\frac 12}{\cal F}^f{}_d\left(\delta^{db}\Xi^a - \delta^{da}\Xi^b\right)
\bigr]\Bigr\}\biggr\}\Biggr\} + {\frac c2}\,\Xi^a \nonumber\\
&& - \,{\frac{ec^2}{2}}\left\{\frac{1}{\epsilon'},V^2{\mathfrak B}^a\right\}
- {\frac {2V}\hbar}{\mathcal{M}}^a\nonumber\\
&& +\,{\frac{c^2}{2\hbar}}\biggl\{\frac{1}{{\cal T}},\bigl\{\{\delta^{ab}{\cal F}^d{}_b{\cal F}^e{}_c
V{\mathcal{M}}^c,\pi_d\},\pi_e\bigr\}\biggr\}.\label{finalOmega}
\end{eqnarray}

The terms on the right-hand side of Eq. (\ref{spinmeq}) contain the two different matrices, $\bm\Sigma$ and $\bm\Pi$, which is related to the fact that $\bm\Omega_{(1)}$ is linear in components of the momentum operator, whereas $\bm\Omega_{(2)}$ depends on the even number of $\pi_a$. As one can notice, the momentum operator enters both vectors $\bm\Omega_{(1)}$ and $\bm\Omega_{(2)}$ only in the combination ${\cal F}^b{}_a \pi_b$. We demonstrate below, see eq. (\ref{Fpv}), that the velocity operator is equal to $\widehat{v}_a = \beta c^2{\cal F}^b{}_a\pi_b/\epsilon'$ and thus it is proportional to $\beta$. As a result, we obtain an additional $\beta$ factor in the operator $\bm\Omega_{(1)}$, and hence both terms on the right-hand side of (\ref{spinmeq}) have the same structure, when everything is rewritten in terms of the velocity operator $\widehat{\bm{v}}$.  It is also worthwhile to notice that only upper part of $\beta$ (proportional the unit $2\times 2$ matrix) is relevant in the FW representation. Therefore, the presence of $\beta$ does not lead to any physical effects (unless antiparticles are considered, when a special analysis is needed). 

We are now in a position to derive the corresponding semiclassical expressions from the results above by evaluating all anticommutators and neglecting the powers of $\hbar$ higher than 1 (for a general discussion of the classical limit of relativistic quantum mechanics, see Ref. \cite{JINRLett12}). The equations (\ref{spinmeq})-(\ref{finalOmega}) then yield the following explicit semiclassical equations describing the precession of the average spin ${\bm s}$ vector (as before, vector product is defined by $\{\bm A\times\bm B\}_a = \epsilon_{abc}A^bB^c$):
\begin{eqnarray}
{\frac {d{\bm s}}{dt}} &=& \bm \Omega\times{\bm s}
= (\bm \Omega_{(1)}+\bm \Omega_{(2)})\times{\bm s},\label{dots}\\
\Omega^a_{(1)} &=& {\frac {c^2}{\epsilon'}}{\cal F}^d{}_c \pi_d\biggl[{\frac 12}
{\Upsilon}\delta^{ac} - \epsilon^{abe}V{\cal C}_{be}{}^c + {\frac {\epsilon'}
{\epsilon' + mc^2V}}\epsilon^{abc}W^e{}_{\widehat{b}}\,\partial_eV\nonumber\\
&& +\,{\frac{eV^2}{\epsilon' + mc^2V}}\,\epsilon^{acb}{\mathfrak E}_b 
- {\frac{2V}{c\hbar}}\epsilon^{acb}{\mathcal{P}}_b\biggr],\label{FO}\\
\Omega^a_{(2)} &=& {\frac c2}\,\Xi^a - {\frac {c^3}{\epsilon'(\epsilon'+mc^2V)}}
\,\epsilon^{abc}{\cal Q}_{(bd)}\delta^{dn}{\cal F}^k{}_n\pi_k{\cal F}^l{}_c\pi_l \nonumber\\
&& -\,{\frac{ec^2V^2}{\epsilon'}}{\mathfrak B}^a + {\frac {2V}\hbar}\left[- {\mathcal{M}}^a
+ {\frac{c^2}{\epsilon'(\epsilon' + mc^2V)}}\delta^{an}{\cal F}^c{}_n\pi_c{\cal F}^d{}_b
\pi_d{\mathcal{M}}^b\right].\label{finalOmegase}
\end{eqnarray}
Here, in the semiclassical limit, we have
\begin{equation}
\epsilon' = \sqrt{m^2c^4V^2 + c^2\delta^{cd}{\cal F}^a{}_c\,{\cal F}^b{}_d\,\pi_a\,\pi_b\,}.\label{eQ}
\end{equation}

Substituting the results obtained into the FW Hamiltonian (\ref{eqFW}), we can recast the latter into a compact and transparent form:
\begin{eqnarray}
{\cal H}_{FW} = \beta\epsilon' + e\Phi + \frac c2\left(\bm K\!\cdot\!\bm\pi
+ \bm \pi\cdot\!\bm K\right) +\frac\hbar2\bm\Pi\cdot\bm\Omega_{(1)}+
\frac\hbar2\bm\Sigma\cdot\bm\Omega_{(2)}.\label{Hamlt}
\end{eqnarray}

Making use of (\ref{Hamlt}), we get the velocity operator in the semiclassical approximation:
\begin{equation}
{\frac {dx^a}{dt}} = \frac{i}{\hbar}[{\cal H}_{FW},x^a] = \beta\,{\frac
{\partial\epsilon'}{\partial \pi_a}} + cK^a
= \beta\,{\frac {c^2}{\epsilon'}}\,{\cal F}^a{}_b\delta^{bc}{\cal F}^d{}_c
\pi_d + cK^a.\label{velocity}
\end{equation}
Let us compare this expression with the relation between the holonomic and anholonomic components of particle's velocity. The anholonomic components of the 4-velocity are conveniently parametrized by the spatial 3-velocity $\widehat{v}^a$ ($a = 1,2,3$) as
\begin{equation}\label{U}
U^\alpha = \left(\begin{array}{c}\gamma \\ \gamma \widehat{v}^a\end{array}\right),
\end{equation}
where $\gamma = (1 - \widehat{v}^2/c^2)^{-1/2}$ is the Lorentz factor ($\widehat{v}^2 = \delta_{ab}\widehat{v}^a\widehat{v}^b$). As a result, we have for the components of the holonomic velocity
\begin{eqnarray}
U^a &=& {\frac {dx^a}{d\tau}} = e^a_\alpha U^\alpha = {\frac \gamma V}
(cK^a + VW^a{}_{\widehat b}\,\widehat{v}^{b}),\label{Ua}\\
U^0 &=& {\frac {dt}{d\tau}} = e^0_\alpha U^\alpha = {\frac \gamma V},\label{U0}
\end{eqnarray}
and hence 
\begin{equation}
{\frac {dx^a}{dt}} = {\frac {U^a}{U^0}} = {\cal F}^a{}_b\,\widehat{v}^b + cK^a.\label{va}
\end{equation}
Comparing this equation with (\ref{velocity}), we can thus identify the velocity operator in the Schwinger frame (\ref{coframe}) with
\begin{equation}
\beta\,{\frac {c^2}{\epsilon'}}\,{\cal F}^b{}_a\pi_b = \widehat{v}_a.\label{Fpv}
\end{equation}
This obviously yields $\delta^{cd}{\cal F}^a{}_c\,{\cal F}^b{}_d \pi_a\pi_b = (\epsilon')^2\widehat{v}^2/c^2$, and by making use of this in (\ref{eQ}), we find $(\epsilon')^2 = m^2c^4V^2 + (\epsilon')^2\widehat{v}^2/c^2$, and consequently: 
\begin{equation}
\epsilon' = \gamma\,mc^2\,V.\label{eV}
\end{equation}
The two equations (\ref{Fpv}) and (\ref{eV}) are decisive for establishing the complete agreement of the quantum and classical dynamics of spin. Namely, from (\ref{Fpv}) and (\ref{eV}) we derive
\begin{equation}
{\frac {\epsilon'}{\epsilon' + mc^2V}} = {\frac \gamma {1 + \gamma}},\qquad 
{\frac {c^3}{\epsilon'(\epsilon' + mc^2V)}}\,{\cal F}^b{}_a\pi_b{\cal F}^d{}_c\pi_d 
= {\frac \gamma {1 + \gamma}}\,{\frac {\widehat{v}_a\widehat{v}_c}{c}},\label{factors}
\end{equation}
which we will use later in the discussion of the quantum-classical correspondence.

\section{Dynamics of classical spin}\label{CS}

In this section we briefly revisit the classical theory of spin in arbitrary external fields.  

The motion of a spinning test particle is characterised by its 4-velocity $U^\alpha$ and by the 4-vector of spin $S^\alpha$ which satisfy the normalization, $U_{\alpha}U^\alpha = c^2$, and the orthogonality, $S_{\alpha}U^\alpha = 0$, conditions. By neglecting the second order spin effects \cite{chicone,dinesh}, which is sufficient in the present study, the dynamical equations for these variables can be written, quite generally, in the form 
\begin{eqnarray}
{\frac {dU^\alpha}{d\tau}} &=& {\cal F}^\alpha,\label{dotU}\\
{\frac {dS^\alpha}{d\tau}} &=& \Phi^\alpha{}_\beta S^\beta.\label{dotS}
\end{eqnarray}
The external fields of various physical nature (electromagnetic, gravitational, scalar, etc.) determine the  forces ${\cal F}^\alpha$ acting on a particle, as well as the spin transport matrix $\Phi^\alpha{}_\beta$ that affects the spin. Normalization and orthogonality of the velocity and spin vectors impose the conditions on the right-hand sides of (\ref{dotU}),(\ref{dotS}):
\begin{equation}
U_\alpha {\cal F}^\alpha = 0,\qquad U_\alpha\Phi^\alpha{}_\beta S^\beta =
-\,S_\alpha {\cal F}^\alpha.\label{cc}
\end{equation}
The spin transport matrix is supposed to be skew-symmetric, $\Phi_{\alpha\beta}=-\Phi_{\beta\alpha}$, which automatically guarantees $S_\alpha S^\alpha =$const.

When the particle is at rest, its spatial 3-velocity vanishes $\widehat{v}^a =0$ and the 4-velocity (\ref{U}) reduces to
\begin{equation}\label{Ur}
u^\alpha = \delta^\alpha_0 = \left(\begin{array}{c}1 \\ 0\end{array}\right).
\end{equation}
The 4-velocity $U^\alpha$ in the laboratory frame (\ref{U}) is related to the rest-frame value (\ref{Ur}) by means of the local Lorentz transformation $U^\alpha = \Lambda^\alpha{}_\beta u^\beta$ where
\begin{equation}\label{Lambda}
\Lambda^\alpha{}_\beta = \left(\begin{array}{c|c}\gamma & \gamma \widehat{v}_b/c^2 \\
\hline \gamma\widehat{v}^a & \delta^a_b + (\gamma - 1)
\widehat{v}^a\widehat{v}_b/\widehat{v}^2\end{array}\right).
\end{equation}

Substituting (\ref{U}) into the orthogonality relation $S_{\alpha}U^\alpha = 0$, we find that the zeroth component of the spin 4-vector is expressed in terms of the 3 spatial components:
\begin{equation}
S^0 = {\frac 1{c^2}}\,\widehat{v}_a S^a.\label{S0}
\end{equation}
The laboratory-frame components of the vector $S^\alpha$ do not describe the physical spin of a particle: We have to recall that spin, as the ``internal angular momentum'' of a particle, is determined with respect to the rest reference frame. We denote this physical spin by $s^\alpha$ (in general, the lower case letters will be used for any other objects in the rest frame). Since the rest frame ($U^\alpha\longrightarrow u^\alpha$) is obtained with the help of the Lorentz transformation (\ref{Lambda}), we have $S^\alpha = \Lambda^\alpha{}_\beta s^\beta$. Inverting this, we find the relation between the physical spin and the 4-vector in the laboratory frame:
\begin{equation}\label{sa}
s^\alpha = (\Lambda^{-1})^\alpha{}_\beta S^\beta =
\left(\begin{array}{c|c}\gamma & -\gamma \widehat{v}_b/c^2 \\ \hline -\gamma
\widehat{v}^a & \delta^a_b + (\gamma - 1)\widehat{v}^a\widehat{v}_b/\widehat{v}^2
\end{array}\right)\left(\begin{array}{c} S^0 \\
S^b\end{array}\right) = \left(\begin{array}{c}0 \\
s^a\end{array}\right).
\end{equation}
Using (\ref{S0}) and the identity $(\gamma - 1)c^2/\widehat{v}^2 \equiv\gamma^2/(\gamma + 1)$, we find explicitly
\begin{equation}
s^a = S^a - {\frac {\gamma}{\gamma + 1}}\,{\frac {\widehat{v}^a\widehat{v}_b}{c^2}}\,S^b.
\end{equation}
Substituting $S^\alpha = \Lambda^\alpha{}_\beta s^\beta$ into (\ref{dotS}), we derive the dynamical equation for the {\it physical spin}:
\begin{equation}
{\frac {ds^\alpha}{d\tau}} = \Omega^\alpha{}_\beta
s^\beta.\label{dsdt}
\end{equation}
Here we introduced
\begin{equation}\label{Omab1}
\Omega^\alpha{}_\beta = \phi^\alpha{}_\beta + \omega^\alpha{}_\beta\,,
\end{equation}
where $\phi^\alpha{}_\beta = (\Lambda^{-1})^\alpha{}_\gamma\Phi^\gamma{}_\delta\Lambda^\delta{}_\beta$ is the rest-frame value of the spin transport matrix $\Phi^\alpha{}_\beta$, and
\begin{equation}
\omega^\alpha{}_\beta := -\,(\Lambda^{-1})^\alpha{}_\gamma{\frac d
{d\tau}} \Lambda^\gamma{}_\beta.\label{Thomas}
\end{equation}
After substituting (\ref{Lambda}) into (\ref{Thomas}), the simple matrix algebra yields 
\begin{equation}
\omega^\alpha{}_\beta = \left(\begin{array}{c|c}0 & -f_b/c^2 \\
\hline -f^a & \omega^a{}_b\end{array}\right),\qquad \omega^a{}_b =
{\frac {\gamma^2}{\gamma + 1}}\left({\frac {\widehat{v}^a}{c^2}}{\frac {d\widehat{v}_b}{d\tau}}
- {\frac {\widehat{v}_b}{c^2}}{\frac {d\widehat{v}^a}{d\tau}}\right).\label{Thomas1}
\end{equation}
Here the rest-frame components $f^\alpha = (\Lambda^{-1})^\alpha{}_\beta {\cal F}^\beta$ of the force 4-vector read
\begin{equation}
f^0 = 0,\qquad f^a = {\cal F}^a - {\frac {\gamma}{\gamma + 1}}
\,{\frac {\widehat{v}^a\widehat{v}_b}{c^2}}\,{\cal F}^b,
\end{equation}
and we used (\ref{dotU}) to obtain the off-diagonal entries in (\ref{Thomas1}).

The formula (\ref{Thomas}) provides perhaps the most straightforward derivation of the {\it Thomas precession}, revealed explicitly in (\ref{Thomas1}). This subject has been recently discussed in great detail in Ref. \cite{PRDThomaspre}.

Computation of the rest-frame components of the spin transport matrix
\begin{equation}
\phi^\alpha{}_\beta = \left(\begin{array}{c|c}0 & \phi^0{}_b \\
\hline \phi^a{}_0 & \phi^a{}_b\end{array}\right)\label{phirest}
\end{equation}
is straightforward: one just needs to evaluate the product of the three matrices, $\phi^\alpha{}_\beta = (\Lambda^{-1})^\alpha{}_\gamma\Phi^\gamma{}_\delta\Lambda^\delta{}_\beta$. The result reads $\phi^0{}_b = \delta_{ab} \phi^a{}_0/c^2$, and
\begin{eqnarray}
\phi^a{}_0 &=& \gamma\left(\Phi^a{}_0 - {\frac {\gamma}{\gamma +
1}}\,{\frac {\widehat{v}^a\widehat{v}_b}{c^2}}\,\Phi^b{}_0 +
\Phi^a{}_b\widehat{v}^b\right),\label{phi0a}\\
\phi^a{}_b &=& \Phi^a{}_b + {\frac 1 {c^2}}\left(\varphi^a\widehat{v}_b
- \varphi_b\widehat{v}^a\right),\label{phiab}\\
\varphi^a &=& \gamma\left(\Phi^a{}_0 + {\frac {\gamma}{\gamma +
1}} \,\Phi^a{}_b\widehat{v}^b\right).\label{phia}
\end{eqnarray}

The physical spin is characterized by the three nontrivial spatial components, (\ref{sa}), and one can prove that the 0-th component of (\ref{dsdt}) vanishes identically (this is equivalent to the second compatibility condition (\ref{cc})). As a result, the dynamical equation for the spin (\ref{dsdt}) reduces to the 3-vector form
\begin{equation}
{\frac {ds^a}{d\tau}} = \Omega^a{}_b s^b,\qquad {\rm or}\qquad
{\frac {d{\bm s}}{d\tau}} = {\bm \Omega}\times{\bm s}.\label{ds1}
\end{equation}
Here the components of the 3-vectors are introduced by ${\bm s} = \{s^a\}$ and ${\bm \Omega} = \left\{-\,{\frac 12}\epsilon^{abc}\Omega_{bc}\right\}$. Recalling (\ref{Omab1}), we find the angular velocity of the spin precession
\begin{equation}
{\bm \Omega} = {\bm \phi} + {\bm \omega},\label{Omab2}
\end{equation}
where ${\bm \phi} = \left\{-\,{\frac 12}\epsilon^{abc}\phi_{bc}\right\}$
and ${\bm \omega} = \left\{-{\,\frac 12}\epsilon^{abc}\omega_{bc}\right\}$

The new general equations (\ref{Thomas})-(\ref{ds1}) are valid for a spinning particle interacting with any external fields. The actual dynamics of the physical spin depends on the forces which act on the particle and on the law of the spin transport.

\section{Classical vs. quantum spinning particles}\label{CSQ}

In the previous section, we have developed a general formalism for the discussion of the dynamics of a classical spinning particle in arbitrary external fields. Now we specialize to the case of the motion of a particle in electromagnetic and gravitational (inertial) fields.

The dynamics of a relativistic particle with mass $m$, electric charge $e$ and dipole moments $\mu', \delta'$ in the gravitational and electromagnetic fields is described by \cite{OSTalkPRD}
\begin{eqnarray}
{\frac {DU^\alpha}{d\tau}} &=& {\frac {dU^\alpha}{d\tau}} + U^i\Gamma_{i\beta}{}^\alpha U^\beta
= -\,{\frac em}\,g^{\alpha\beta}F_{\beta\gamma}U^\gamma,\label{DUG}\\
{\frac {DS^\alpha}{d\tau}} &=& {\frac {dS^\alpha}{d\tau}} + U^i\Gamma_{i\beta} {}^\alpha S^\beta
=  -\,{\frac em}\,g^{\alpha\beta}F_{\beta\gamma}S^\gamma \nonumber\\
&& -\,{\frac 2\hbar}\left[M^\alpha{}_\beta + {\frac {1}{c^a}}\left(M_{\beta\gamma}U^\alpha
U^\gamma - M^{\alpha\gamma}U_\beta U_\gamma\right)\right]S^\beta.\label{DSG}
\end{eqnarray}
Here we introduced
\begin{equation}
M_{\alpha\beta} = \mu' F_{\alpha\beta} + c\delta'\,G_{\alpha\beta},\label{Mab}
\end{equation}
with the components
\begin{equation}
M_{\hat{0}\hat{a}} = c{\mathcal P}_a,\qquad
M_{\hat{a}\hat{b}} = \epsilon_{abc}{\mathcal M}^c,\label{MP}
\end{equation}
where the 3-vectors $\bm{\mathcal M}$ and $\bm{\mathcal P}$ were defined in (\ref{MaPa}).

Following \cite{chicone,dinesh}, we describe the motion of classical spinning particles by the system (\ref{DUG})-(\ref{DSG}) in which the second order spin effects are neglected. This is consistent with the construction of the quantum FW picture up to the first order in the Planck constant. The analysis of higher-order spin effects in the general Mathisson-Papapetrou model of the classical spinning particle is a more difficult problem which is extensively studied in the literature, see \cite{plyatsko1,plyatsko2,Dirk1,Dirk2,Damb1,Damb2,Lukes1,Lukes2,Hojman,Duval} and the references therein. 

In accordance with the general scheme of Sec.~\ref{CS}, we write the explicit force and the spin transport matrix for the system (\ref{DUG})-(\ref{DSG}):
\begin{eqnarray}
{\cal F}^\alpha &=& -\,U^i\Gamma_{i\beta}{}^\alpha U^\beta
- {\frac {e}{m}}\,F^\alpha{}_\beta\,U^\beta,\label{Fe}\\ \label{Pe}
\Phi^\alpha{}_\beta &=& -\,U^i\Gamma_{i\beta}{}^\alpha - {\frac {e}{m}}
\,F^\alpha{}_\beta - {\frac 2\hbar}\left[M^\alpha{}_\beta + {\frac {1} {c^2}}
\left(U^\alpha M_{\beta\gamma}U^\gamma - U_\beta M^{\alpha\gamma} U_\gamma\right)\right].
\end{eqnarray}
One can check that the compatibility conditions (\ref{cc}) are satisfied.

Substituting (\ref{Fe}), (\ref{Pe}) and (\ref{U}) into
(\ref{phi0a})-(\ref{phia}) and (\ref{Thomas1}), we derive
\begin{equation}\label{phiomsum}
\bm{\phi} = {\stackrel {(e)}{\bm{\phi}}} + {\stackrel {(g)}{\bm{\phi}}},\qquad
\bm{\omega} = {\stackrel {(e)}{\bm{\omega}}} + {\stackrel {(g)}{\bm{\omega}}}.
\end{equation}
The electromagnetic field contributions read
\begin{eqnarray}
{\stackrel {(e)}{\bm{\phi}}} &=& {\frac {e}{m}}\,\gamma\left[- \bm{\mathfrak B}
+ {\frac {\gamma} {\gamma + 1}}\,{\frac {\widehat{\bm v}(\widehat{\bm v}\cdot
\bm{\mathfrak B})}{c^2}} + {\frac {\widehat{\bm v}\times \bm{\mathfrak E}}{c^2}}
\right]\nonumber\\
&& + \,{\frac {2\gamma}\hbar}\left[- \bm{\mathcal M} + {\frac {\gamma} {\gamma + 1}}
\,{\frac {\widehat{\bm v}(\widehat{\bm v}\cdot\bm{\mathcal M})}{c^2}}
- {\frac {\widehat{\bm v}\times \bm{\mathcal P}}{c}}\right],\label{phiE}\\
{\stackrel {(e)}{\bm{\omega}}} &=& {\frac {e}{m}}\,(\gamma  - 1)\left[
\bm{\mathfrak B} - {\frac {\widehat{\bm v} (\widehat{\bm v}\cdot\bm{\mathfrak B})
+ \widehat{\bm v}\times\bm{\mathfrak E}}{\widehat{v}^2}}\right],\label{omegaE}
\end{eqnarray}
and the gravitational field contributions are
\begin{eqnarray}\label{phiG}
{\stackrel {(g)}{\bm{\phi}}}{}_a &=& U^i\epsilon_{abc}\left[{\frac 12}
\Gamma_i{}^{cb} + {\frac {\gamma^2}{\gamma + 1}} \,{\frac {\widehat{v}_d}{c^2}}
\,\Gamma_{id}{}^b\widehat{v}^c + {\frac {\gamma}{c^2}}\,\Gamma_{i\widehat{0}}{}^b
\widehat{v}^c\right],\\
{\stackrel {(g)}{\bm{\omega}}}{}_a &=& -\,{\frac {\gamma^2}{\gamma + 1}}
U^i\epsilon_{abc}\left[{\frac {\widehat{v}_d} {c^2}}\,\Gamma_{id}{}^b\widehat{v}^c
+ {\frac 1 {c^2}}\,\Gamma_{i\widehat{0}}{}^b\widehat{v}^c\right].\label{omegaG}
\end{eqnarray}

The precession of the physical spin is the sum (\ref{Omab2}). The result reads
\begin{equation}\label{Omtot}
\bm{\Omega} = {\stackrel {(e)}{\bm{\Omega}}} + {\stackrel {(g)}{\bm{\Omega}}},
\end{equation}
where we find for the electromagnetic ${\stackrel {(e)}{\bm{\Omega}}} =
{\stackrel {(e)}{\bm{\phi}}} + {\stackrel {(e)}{\bm{\omega}}}$ and the gravitational
${\stackrel {(g)}{\bm{\Omega}}} = {\stackrel {(g)}{\bm{\phi}}} +
{\stackrel {(g)}{\bm{\omega}}}$  parts, respectively:
\begin{eqnarray}
{\stackrel {(e)}{\bm{\Omega}}} &=& {\frac {e}{m}}\left[- \bm{\mathfrak B}
+ {\frac {\gamma} {\gamma + 1}}{\frac {\widehat{\bm v}\times\bm{\mathfrak E}}{c^2}}
\right] + \,{\frac {2\gamma}\hbar}\left[ - \bm{\mathcal M} + {\frac {\gamma}
{\gamma + 1}}\,{\frac {\widehat{\bm v}(\widehat{\bm v}\cdot\bm{\mathcal M})}{c^2}}
- {\frac {\widehat{\bm v}\times \bm{\mathcal P}}{c}}\right],\label{OmegaE}\\
{\stackrel {(g)}{\bm{\Omega}}}{}_a &=& \epsilon_{abc}\,U^i\left[{\frac 12}
\Gamma_i{}^{cb} + {\frac {\gamma}{\gamma + 1}}\,\Gamma_{i\widehat{0}}{}^b
\widehat{v}^c/c^2\right].\label{OmegaG}
\end{eqnarray}
The exact formula (\ref{OmegaG}) can also be used in flat spacetimes for noninertial reference frames and curvilinear coordinates (see the relevant discussion in Ref. \cite{JINRLettCylr} and in the references therein), since the connection $\Gamma_{i\beta}{}^\alpha$ contains information about both gravitational and inertial effects. Substituting (\ref{connection1}) and (\ref{connection2}), we can recast (\ref{OmegaG}) into
\begin{equation}
{\stackrel {(g)}{\bm{\Omega}}} = -\bm{\mathcal{B}} + {\frac {\gamma}{\gamma +
1}}\,{\frac {\widehat{\bm{v}} \times\bm{\mathcal{E}}} {c^2}},\label{omgem}
\end{equation}
where the generalized gravitoelectric and gravitomagnetic fields are defined by
\begin{eqnarray}\label{ge}
{\mathcal E}^a &=& {\frac {\gamma}V}\delta^{ac}\left(c{\cal Q}_{(\widehat{c}\widehat{b})}
\widehat{v}^b - c^2\,W^b{}_{\widehat{c}}\,\partial_bV\right),\\
{\mathcal B}^a &=& {\frac {\gamma}V}\left(-\,{\frac c2}\,{\Xi}^a - {\frac 12}\Upsilon
\,\widehat{v}^a + \epsilon^{abc}V{\cal C}_{bc}{}^d\widehat{v}_d\right).\label{gm}
\end{eqnarray} 
For the first discussion of these fields see Pomeransky and Khriplovich \cite{PK} (cf. also Refs. \cite{OST,OSTRONG,ostgrav,PRDThomaspre} for more details). The remarkable similarity of (\ref{omgem}) and of the first term in (\ref{OmegaE}) suggests introducing the effective magnetic and electric fields
\begin{eqnarray}
\bm{\mathfrak B}_{\rm eff} &=& \bm{\mathfrak B} + {\frac me}\,\bm{\mathcal B},\label{Beff}\\
\bm{\mathfrak E}_{\rm eff} &=& \bm{\mathfrak E} + {\frac me}\,\bm{\mathcal E}.\label{Eeff}
\end{eqnarray}
Accordingly, the general precession velocity (\ref{Omtot}) is rewritten as
\begin{equation}
\bm{\Omega} = {\frac {e}{m}}\left[- \bm{\mathfrak B}_{\rm eff} +
{\frac {\gamma} {\gamma + 1}}{\frac {\widehat{\bm v}\times\bm{\mathfrak E}_{\rm eff}}{c^2}}
\right] + \,{\frac {2\gamma}\hbar}\left[ - \bm{\mathcal M} + {\frac {\gamma}
{\gamma + 1}}\,{\frac {\widehat{\bm v}(\widehat{\bm v}\cdot\bm{\mathcal M})}{c^2}}
- {\frac {\widehat{\bm v}\times \bm{\mathcal P}}{c}}\right].\label{OmegaT}
\end{equation}
This generalizes our findings in \cite{OSTalkPRD}.

Making use of Eqs. (\ref{Fpv})-(\ref{factors}) we thus finally prove the complete agreement between the classical limit of the quantum-mechanical dynamics (\ref{dots})-(\ref{finalOmegase}) and the corresponding equations of motion of classical spin (\ref{ds1}) and (\ref{OmegaT}) in the most general case of arbitrary gravitational (inertial) and electromagnetic field acting on the particle. Note that one should use (\ref{U0}) to relate the derivatives with respect to the proper and the coordinate time, ${\frac {d}{d\tau}} = {\frac {\gamma}{V}}{\frac {d}{dt}}$. The demonstration of consistency of the quantum and classical spin dynamics is the main result of our paper. 

\section{Example: spinning particle in noninertial frame}\label{Example}

The formalism developed and the results obtained are completely general. As a first application, we consider the motion of a spinning particle in a noninertial system which is described by the metric of the flat spacetime of an accelerating and rotating observer \cite{HehlNi}
\begin{equation}
V = 1 + {\frac {{\bm a}\cdot{\bm r}}{c^2}},\qquad W^{\widehat a}{}_b = \delta^a_b,
\qquad K^a =-\,{\frac 1c}\,(\bm\omega\times\bm r)^a.\label{VWni}
\end{equation}
Here ${\bm a}$ describes acceleration of the observer and $\bm\omega$ is an angular velocity of a noninertial reference system; both are independent of the spatial coordinates. Acceleration and rotation have a clear physical meaning as the gravitoelectric (\ref{ge}) and gravitomagnetic (\ref{gm}) fields: 
\begin{equation}\label{gemHN}
\bm{\mathcal E} = -\,{\frac \gamma V}\,\bm{a},\qquad \bm{\mathcal B} = {\frac \gamma V}\,\bm{\omega}.
\end{equation}

The primary goal of considering this example is to show how the formalism actually works. At the same time, this case is not of a pure academic interest, because of its importance for the discussion of the validity of the equivalence principle for quantum-mechanical systems. 

In the flat spacetime of a general noninertial system (\ref{VWni}), the relations between the holonomic and anholonomic electric and magnetic fields (\ref{EE}) and (\ref{BB}) reduce to
\begin{equation} 
\bm{\mathfrak{E}} = {\frac 1V}\,(\bm{E} + c\bm{K}\times\bm{B}),\qquad
\bm{\mathfrak{B}} = \bm{B}.\label{EBni}
\end{equation}
At the same time, for the anholonomic velocity (\ref{va}) we recover the familiar classic formula
\begin{equation}
\widehat{\bm{v}} = \dot{\bm{r}} - c\bm{K} = \bm{v} + \bm\omega\times\bm r.\label{vani}
\end{equation} 

For simplicity, we confine our attention to the minimally coupled particles for which the AMM and EDM are absent: $\mu' = 0$  and $\delta' = 0$. The corresponding semiclassical FW Hamiltonian then reads
\begin{eqnarray}
{\cal H}_{FW} &=& \frac\beta2\left\{\left(1+\frac{\bm a\cdot\bm r}{c^2}\right), 
\sqrt{m^2c^4+c^2\bm{\pi}^2}\right\} - \bm\omega\cdot({\bm r}\times{\bm \pi})\nonumber\\
&& +\,\frac\hbar2\bm\Pi\cdot\bm\Omega^{(1)} + \frac\hbar2\bm\Sigma\cdot\bm\Omega^{(2)},\label{Hamltni}
\end{eqnarray}
where the angular velocity operators are given by
\begin{eqnarray}\label{Om1HN}
\bm\Omega^{(1)} = \frac{\left(\bm{a} - {\frac {eV}{m\gamma}}\bm{\mathfrak E}\right)
\times\bm\pi}{mc^2(\gamma+1)},\qquad \bm\Omega^{(2)} = -\,\bm\omega 
- {\frac {eV}{m\gamma}}\bm{\mathfrak B},\label{Om2HN}
\end{eqnarray}
with $\gamma=\frac{\sqrt{m^2c^4+c^2\bm\pi^2}}{mc^2}$. In the semiclassical limit we have a remarkable relation:
\begin{equation}
\bm{\pi} = m\widehat{\bm{v}}\gamma = {\frac {m\widehat{\bm{v}}}{\sqrt{1 - {\frac {\widehat{v}^2}{c^2}}}}}.\label{piv}
\end{equation}
This relativistic formula shows that the canonical momentum is directed along the anholonomic velocity (\ref{vani}), and not along the coordinate velocity $\bm{v} = \dot{\bm{r}}$. 

In high-energy experimental physics, it is more convenient to describe particle's motion with respect to detectors. In this case, one should subtract the angular velocity of the particle revolution from the precession angular velocity $\bm\Omega = \bm\Omega^{(1)} + \bm\Omega^{(2)}$ and use the cylindrical or Frenet-Serret coordinate systems. The spin dynamics in these coordinate systems looks differently, for the relevant discussion see Ref. \cite{JINRLettCylr} and references therein. When the Frenet-Serret coordinates are used, one needs to find the angular velocity of rotation of the unit vector $\widehat{\bm{N}} = {\frac {\bm{\pi}}{\pi}} = {\frac {\widehat{\bm{v}}}{\widehat{v}}}$. This unit vector determines the direction of the motion. Making use of particle's equations of motion, obtained from the FW Hamiltonian (\ref{Hamltni}), we derive
\begin{equation}
{\frac {d\widehat{\bm{N}}}{dt}} = \widehat{\bm{O}}\times\widehat{\bm{N}},\label{dNdta}
\end{equation}
where we find that the rotation of the direction is described by
\begin{eqnarray}
\widehat{\bm{O}} = -\,\bm{\omega} - {\frac {e}{m\gamma}}\bm{\mathfrak B} 
- {\frac {\widehat{\bm{v}}}{\widehat{v}{}^2}}\times\left(\bm{a} - {\frac {e}{m\gamma}}
\bm{\mathfrak E}\right).\label{Oa}
\end{eqnarray}
As a result, the precession of spin in the Frenet-Serret coordinates is governed by 
\begin{eqnarray}
\bm{\Omega}^{FS} = \bm{\Omega} - \widehat{\bm{O}} 
= {\frac 1{\gamma^2 - 1}}\,{\frac {\widehat{\bm{v}}}{c^2}}\times
\left(\bm{a} - {\frac {e}{m\gamma}}\bm{\mathfrak E}\right).\label{OMFS}
\end{eqnarray}
Quite remarkably, the explicit contribution of the rotation $\bm{\omega}$ and the magnetic field $\bm{\mathfrak{B}}$ disappeared from $\bm{\Omega}^{FS}$. Nevertheless, the inertial effects of rotation and magnetic effects are still present implicitly in the anholonomic variables (\ref{EBni}) and (\ref{vani}).

This example clarifies the physical meaning of the gravitoelectric and gravitomagnetic fields and shows their natural similarity with the electric and magnetic fields in the way they affect the dynamics of spin.

\section{Spin in the field of a gravitational wave}\label{SG}

Let us apply the general results obtained above to the study of the spin dynamics of a fermion particle in the gravitational wave in the presence of the electromagnetic field. Previously, this was analyzed in \cite{gos,Quach1,Quach2}. Gravitational waves are the phenomena of fundamental importance in physics, and quite remarkably the theoretical research \cite{griff,vdz,exact} of this subject has been recently supported by the first experimental evidence \cite{Abbott1,Abbott2,Abbott3}. An informative historic overview can be found in \cite{flan,schutz,CNN}. 

\subsection{Gravitational wave}\label{GW}

It will be convenient to use the general framework which allows to deal both with the exact gravitational wave solutions of Einstein's theory and with the weak approximate wave configurations. 

We choose the spacetime local coordinates as $t, x^1 = x, x^2 = y, x^3 = z$, and consider a general plane-fronted gravitational wave with the spacetime interval
\begin{equation}
ds^2 = c^2dt^2  - \underline{g}{}_{AB}\,dx^Adx^B - dz^2 + U(cdt - dz)^2.\label{dsw}
\end{equation}
The metric coefficients $\underline{g}{}_{AB} = \underline{g}{}_{AB}(\sigma)$ and $U = U(\sigma, x^A)$ are the functions of the coordinates $x^A = \{x, y\}$ (with $A,B = 1,2$), parameterizing the wave front, and the parameter $\sigma = ct - z$ along the ray. From the physical point of view, it would be more correct to treat $\underline{g}{}_{AB} = \underline{g}{}_{AB}(\varphi)$ and $U =  U(\varphi, x^A)$ as the functions of wave's phase $\varphi = \omega(t - z/c) = {\frac \omega c}\sigma$, but we will use the geometric variable $\sigma$ instead, which of course yields completely equivalent results. This configuration describes the plane-fronted gravitational wave with the frequency $\omega$ propagating along the $x^3 = z$ axis.

By introducing a new function $V$ via 
\begin{equation}
U = 1 - {\frac 1{V^2}},\label{UV}
\end{equation}
we write the corresponding Schwinger coframe (\ref{coframe}) as
\begin{equation}
\vartheta^{\widehat 0} = Vcdt,\quad \vartheta^{\widehat A} = h^{\widehat{A}}{}_B dx^B,
\quad \vartheta^{\widehat 3} = {\frac 1V}\left(dz + (V^2 - 1)cdt\right),\label{cofwave}
\end{equation}
where the zweibein $h^{\widehat{A}}{}_B =  h^{\widehat{A}}{}_B(\sigma)$, $\widehat{A}, \widehat{B} = 1,2$, satisfy $h^{\widehat{C}}{}_Ah^{\widehat{D}}{}_B\delta_{\widehat{C}\widehat{D}} = \underline{g}{}_{AB}$. The zweibein is not necessarily diagonal, but one can always choose it so that the $2\times 2$ matrix
\begin{equation}\label{PAB}
\Phi^{\widehat{A}}{}_{\widehat{B}} = h^C{}_{\widehat{B}}\partial_\sigma h^{\widehat{A}}{}_C
\end{equation}
(with the inverse zweibein $h^C{}_{\widehat{B}}$) is symmetric
\begin{equation}\label{Psym}
\Phi_{\widehat{A}\widehat{B}} = \delta_{\widehat{A}\widehat{C}}\Phi^{\widehat{C}}{}_{\widehat{B}}
= \delta_{\widehat{B}\widehat{C}}\Phi^{\widehat{C}}{}_{\widehat{A}} = \Phi_{\widehat{B}\widehat{A}}.
\end{equation}
Recalling the general coframe (\ref{coframe}), we now have explicitly
\begin{eqnarray}
W^{\widehat a}{}_b &=& \left(\begin{array}{cc}h^{\widehat A}{}_B & 0 \\
0 & V^{-1}\end{array}\right),\label{Ww}\\
K^a &=& \left\{ 0, 0, 1 - V^2\right\}.\label{Kw}
\end{eqnarray}
For the gravitational wave (\ref{dsw}), we now have 
\begin{equation}\label{gabGW}
\underline{g}{}_{ab} = \left(\begin{array}{cc}\underline{g}{}_{AB} & 0 \\ 0 & U\end{array}\right),\qquad
\underline{g}{}^{ab} = \left(\begin{array}{cc}\underline{g}{}^{AB} & 0 \\ 0 & U^{-1}\end{array}\right).
\end{equation}
The general plane-fronted gravitational wave (\ref{dsw}) encompasses the two special cases: (i) Rosen-Bondi wave \cite{rosen1937,einrosen,bondi0,bondi1} with $U = 0$, and hence $V = 1$ and $K^a = 0$, and arbitrary $h^{\widehat A}{}_B$; (ii) Brinkmann-Peres wave \cite{Brink1,Brink2,Brink3,peres,pen1,pen2} with $h^{\widehat A}{}_B = \delta^A_B$ and arbitrary $U$.

The phase function $\varphi = \omega(t - z/c)$ gives rise to the physical wave covector $d\varphi$.
However, technically it is more convenient to introduce the geometrical wave covector via
\begin{equation}
k = d\sigma = k_\alpha\vartheta^\alpha.\label{kdf}
\end{equation}
The corresponding (anholonomic) components of the wave covector read
\begin{equation}
k_\alpha = \left(cV, 0, 0, -V\right).\label{ka}
\end{equation}
The wave covector is obviously constant, null, and geodetic
\begin{equation}
dk_\alpha = 0,\qquad k_\alpha k^\alpha = 0,\qquad k\wedge {}^*Dk_\alpha = 0.\label{k2}
\end{equation}

One can straightforwardly find the local Lorentz connection 1-form $\Gamma_\beta{}^\alpha = \Gamma_{i\beta}{}^\alpha\,dx^i$ for the gravitational wave (\ref{dsw}). The nontrivial components are as follows:
\begin{equation}
\Gamma_\alpha{}^{\widehat{A}} = k_\alpha\left(\Phi^{\widehat{A}} + k\,D^{\widehat{A}}U/2\right),
\qquad \Gamma_{\widehat{0}}{}^{\widehat{3}} = -\,{\frac cV}dV.\label{Gw}
\end{equation}
Here we denoted $D^{\widehat{A}}U : = \delta^{\widehat{A}\widehat{B}}e_{\widehat{B}}\rfloor dU$, and 
introduced the 1-form, recall (\ref{PAB}),
\begin{equation}\label{Pa}
\Phi^{\widehat{A}} = \Phi^{\widehat{A}}{}_{\widehat{B}}\vartheta^{\widehat{B}}.
\end{equation}

Computation of the curvature 2-form is straightforward, and the result reads
\begin{equation}
R_\alpha{}^\beta = k\wedge (k_\alpha\,\Omega^\beta - k^\beta\,\Omega_\alpha),\label{curv}
\end{equation}
where the vector-valued 1-form $\Omega^\alpha = \Omega^\alpha{}_\beta \vartheta^\beta$ has the 
components $\Omega^{\widehat{0}} = \Omega^{\widehat{3}} = 0$, and  
\begin{equation}\label{curv1}
\Omega^{\widehat{A}} = \left(\partial_\sigma\Phi^{\widehat{A}}{}_{\widehat{B}} + \Phi^{\widehat{A}}{}_{\widehat{C}}
\Phi^{\widehat{C}}{}_{\widehat{B}} + {\frac 12}D_{\widehat{B}}D^{\widehat{A}}U\right)\vartheta^{\widehat{B}}.
\end{equation}
Accordingly, the tensor $\Omega^\alpha{}_\beta$ has the algebraic properties:
\begin{equation}
k_\alpha\Omega^\alpha{}_\beta = 0,\qquad \Omega_{\alpha\beta} = \Omega_{\beta\alpha}.\label{Op}
\end{equation}
Consequently, the Ricci 1-form is then
\begin{equation}
e_\beta\rfloor R_\alpha{}^\beta = -\,k\,k_\alpha\,\Omega^\beta{}_\beta.\label{Ricw}
\end{equation}
As a result the vacuum Einstein equation $e_\beta\rfloor R_\alpha{}^\beta = 0$ reduces
to the scalar equation
\begin{equation}
\partial_\sigma\Phi^{\widehat{A}}{}_{\widehat{A}} + \Phi^{\widehat{A}}{}_{\widehat{B}}
\Phi^{\widehat{B}}{}_{\widehat{A}} + {\frac 12}\Delta U = 0,\label{Ein2}
\end{equation}
where $\Delta = \delta^{AB}\partial^2_{AB}$ is the Laplace operator on the 2-dimensional wave front surface. Recalling (\ref{PAB}), we thus have the second order ordinary differential equation for the zweibein components $h^{\widehat{A}}{}_B = h^{\widehat{A}}{}_B(\sigma)$ and the function $U(\sigma)$.

A weak gravitational wave is described by $h^{\widehat{A}}{}_B = \delta^{\widehat{A}}{}_B
+ w^{\widehat{A}}{}_B$ and $U = 0$, where the components of the $2\times 2$ matrix
\begin{equation}
\begin{array}{c}
w^{\widehat{A}}{}_B = \left(\begin{array}{cc} w_{\bigoplus} & w_{\bigotimes} \\ 
w_{\bigotimes} & - w_{\bigoplus}\end{array}\right)\end{array}\label{Ein3}
\end{equation}
are two arbitrary $w_{\bigoplus}(\varphi),\, w_{\bigotimes}(\varphi)$ small functions ($w_{\bigoplus}\ll 1,\, w_{\bigotimes}\ll 1$) of the phase $\varphi = \omega(t - z/c)$. They correspond to the two possible polarizations of the wave. It is convenient to introduce the $3\times3$ matrix
\begin{equation}\begin{array}{c}
w^a{}_b = \left(\begin{array}{cc}w^{\widehat A}{}_B & 0
\\ 0 & 1\end{array}\right).
\label{dublv}\end{array}\end{equation} 
We will not make a difference between upper and lower indices of this basic matrix.

Many exact gravitational wave solutions are also known \cite{rosen1937,einrosen,bondi0,bondi1,Brink1,Brink2,Brink3,peres,pen1,pen2,griff,vdz,exact}.

\subsection{Fermion in a gravitational wave}\label{spinGW}

The general Dirac Hamiltonian has the form (\ref{HamiltonDP}). Now we specialize to the case of the gravitational wave and using (\ref{Gw}) we calculate the components of the objects (\ref{Qab}) and (\ref{Cabc}). We will not give $Q_{\widehat{a}\widehat{b}}$ and $C_{\widehat{b}\widehat{c}}{}^{\widehat{a}}$ since they enter the Hamiltonian only in combinations (\ref{AB2}) and (\ref{AB3}). The direct computation yields for the latter:
\begin{eqnarray}
\Upsilon = 0,\qquad \Xi^A = 2\eta^{AB}W^C{}_{\widehat{B}}\partial_CV,\qquad 
\Xi^{\widehat{3}} = 0,\label{UXW}
\end{eqnarray}
where $\eta^{AB} = - \eta^{BA}$ (and $\eta_{AB} = - \eta_{BA}$) is the totally antisymmetric Levi-Civita tensor in 2 dimensions on the wave front surface, normalized by $\eta_{12} = \eta^{12} = 1$. As for $V$ and $\bm K$, they are both nontrivial for nonvanishing $U$, see the explicit formulas (\ref{UV}) and (\ref{Kw}). Accordingly, the Dirac Hamiltonian (\ref{HamiltonDP}) reduces to
\begin{eqnarray}
{\cal H} &=& \beta mc^2V + e\Phi + {\frac c 2}\left(\pi_b\,{\cal F}^b{}_a \alpha^a
+ \alpha^a{\cal F}^b{}_a\pi_b\right) - \beta \left(\bm{\Sigma}\cdot\bm{\mathcal M}
+ i\bm{\alpha}\cdot\bm{\mathcal P}\right) \nonumber\\
&& +\,c\pi_z - {\frac {V^2c\pi_z + c\pi_zV^2}2} + {\frac {\hbar c} 2}
\,\eta^{AB}\,\Sigma_{\widehat A} W^C{}_{\widehat B} \partial_CV.\label{HamiltonDW}
\end{eqnarray}

Here (\ref{AB1}) now reads ${\cal F}^b{}_a = V{W}^b{}_{\widehat a}$, or explicitly ${\cal F}^B{}_A = Vh^B{}_{\widehat{A}}, \,{\cal F}^3{}_3 = V^2$. The anholonomic electric (\ref{EE}) and magnetic (\ref{BB}) fields are
\begin{eqnarray}\label{EEw}
{\mathfrak{E}}{}_A &=& {\frac 1V}{W}^B{}_{\widehat A}\left[E_B - c(1 - V^2)\eta_{BC}B^C\right],
\qquad {\mathfrak{E}}{}_3 = E_3,\\ \label{BBw}
{\mathfrak{B}}{}^A &=& {\frac Vh}\,W^{\widehat A}{}_B\,{B}^B,\qquad {\mathfrak{B}}{}^3 ={\frac 1h}B^3.
\end{eqnarray}
Here $h = \det h^{\widehat A}{}_B$. The constitutive relations (\ref{const1})-(\ref{const2}) reduce to
\begin{eqnarray}
D^A &=& \varepsilon_0(1 - U)\,h\underline{g}{}^{AB}E_B - \lambda_0U\eta^{AB}\underline{g}{}_{BC}
B^C,\qquad D^3 = \varepsilon_0\,h\,E_3,\label{DE3w}\\
H_A &=& {\frac {1 + U}{\mu_0h}}\,\underline{g}{}_{AB}B^B - \lambda_0U\underline{g}{}_{AB}\eta^{BC}E_C,
\qquad H_3 = {\frac 1{\mu_0h}}\,B^3.\label{HB3w}
\end{eqnarray}
It is worthwhile to notice that in the gravitational wave with nontrivial $U$ one observes a natural magnetoelectricity.

\subsection{Foldy-Wouthuysen Hamiltonian and equations of motion}\label{FWw}

From now on, we specialize (without actually loosing generality because the two types of gravitational waves are related by a coordinate transformation) to the Rosen-Bondi waves and put $U = 0$. Hence $V = 1$, and we then also find ${\cal F}^b{}_a = W^b{}_{\widehat{a}}$ (or explicitly ${\cal F}^B{}_A = h^B{}_{\widehat{A}}, \,{\cal F}^3{}_3 = 1$) and $\bm\Xi = 0$ from (\ref{UXW}). As a result, the last line in the Dirac Hamiltonian (\ref{HamiltonDW}) vanishes. 

In the case under consideration, the constituents (\ref{eq7})-(\ref{eqa}) of the general FW Hamiltonian reduce to
\begin{eqnarray}
{\cal H}_{FW}^{(1)} &=& \beta\epsilon' + \frac{\hbar c^2}{8}\left\{{\frac{1}{\epsilon'}},
\epsilon^{cae}\Pi_e \{\pi_b,{\cal F}^d{}_c\partial_d{\cal F}^b{}_a\}\right\},\label{eq7w}\\
{\cal H}_{FW}^{(2)} &=& \frac{\hbar c}{16}\Biggl\{\frac{1}{{\cal T}},\biggl\{\Sigma_a \{\pi_e,
{\cal F}^e{}_b\},\Bigl\{\pi_f,\epsilon^{abc}\dot{\cal F}^f{}_c \Bigr\}\biggr\}\Biggr\},\label{eq7Kw}
\end{eqnarray}
\begin{eqnarray}
{\cal H}_{FW}^{(3)} &=&  e\Phi - {\frac{e\hbar c^2}{4}}\left\{\frac{1}{\epsilon'},\Pi^a
{\mathfrak B}_a\right\}\nonumber\\
&& + \,\frac{e\hbar c^2}{8}\left\{\frac{1}{{\cal T}},\Bigl[\Sigma_a\epsilon^{abc}\left(
\{{\cal F}^d{}_b,\pi_d\}{\mathfrak E}_c - {\mathfrak E}_b\{{\cal F}^d{}_c,\pi_d\}\right)
- 2\hbar{\cal F}^b{}_a\partial_b({\mathfrak E}^a)\Bigr]\right\},\label{eqw73}\\
{\cal H}_{FW}^{(4)} &=&
- \,{\frac c8}\biggl\{\frac{1}{\epsilon'},\Bigl[\Sigma_a\epsilon^{abc}\bigl(
\{{\cal F}^d{}_b,\pi_d\}{\mathcal{P}}_c - {\mathcal{P}}_b\{{\cal F}^d{}_c,\pi_d\}\bigr)
- 2\hbar{\cal F}^b{}_a\partial_b({\mathcal{P}}^a)\Bigr]\biggr\} - \Pi^a{\mathcal{M}}_a \nonumber\\
&& + \,\frac{c^2}{4}\biggl\{\frac{1}{{\cal T}},\Bigl(\Pi^a\bigl\{\{{\cal F}^c{}_a{\cal F}^d{}_b
{\mathcal{M}}^b,\pi_c\},\pi_d\bigr\}+\beta\hbar\left\{{\cal F}^b{}_a\mathcal{J}^a,\pi_b
\right\}\Bigr)\biggr\},\label{eqw74}\\
\epsilon' &=& \sqrt{m^2c^4+\frac{c^2}{4}\delta^{ac}\{\pi_b,{\cal F}^b{}_a\}
\{\pi_d,{\cal F}^d{}_c\}},\qquad {\cal T}=2{\epsilon'}(\epsilon' + mc^2),\nonumber\\
\mathcal{J}^a &=& e^{abc}{\cal F}^d{}_b \partial_d(\mathcal{M}_c) 
+ {\frac {\partial\mathcal{P}^a}{c\partial t}}.\label{eqaw}
\end{eqnarray}

The quantum spin dynamics is described by Eq. (\ref{spinmeq}) where
\begin{eqnarray}
\Omega^a_{(1)} &=& \frac{c^2}{4}\left\{\frac{1}{\epsilon'},\{\pi_e,\epsilon^{abc}
{\cal F}^d{}_b\partial_d{\cal F}^e{}_c \} \right\}\nonumber\\
&& +\,\frac{ec^2}{4}\epsilon^{abc}\left\{\frac{1}{{\cal T}},\left(\{{\cal F}^d{}_b,\pi_d\}
{\mathfrak E}_c - {\mathfrak E}_b\{{\cal F}^d{}_c,\pi_d\}\right) \right\} \nonumber\\
&& -\,{\frac c{4\hbar}}\epsilon^{abc}\biggl\{\frac{1}{\epsilon'},\bigl(\{{\cal F}^d{}_b,\pi_d\}
{\mathcal{P}}_c - {\mathcal{P}}_b\{{\cal F}^d{}_c,\pi_d\}\bigr)\biggr\},\label{eqolw}
\end{eqnarray}
and
\begin{eqnarray}
\Omega^a_{(2)} &=& \frac{c}{8}\Biggl\{\frac{1}{{\cal T}},
\biggl\{ \{\pi_e,{\cal F}^e{}_b\},\Bigl\{\pi_f,\epsilon^{abc}
\dot{\cal F}^f{}_c \Bigr\}\biggr\}\Biggr\}  \nonumber\\
&& - \,{\frac{ec^2}{2}}\left\{\frac{1}{\epsilon'},{\mathfrak B}^a\right\} - {\frac {2}\hbar}
{\mathcal{M}}^a + {\frac{c^2}{2\hbar}}\biggl\{\frac{1}{{\cal T}},\bigl\{\{\delta^{an}
{\cal F}^c{}_n{\cal F}^d{}_b{\mathcal{M}}^b,\pi_c\},\pi_d\bigr\}\biggr\}. \label{finalOmegw}
\end{eqnarray}

Comparing to the previous investigations of the spin interaction of the Dirac fermion with the gravitational wave, it is worthwhile to stress that we have derived the relativistic formulas without using the weak-field approximation. Since the FW Hamiltonian (\ref{eqFW}) is exact in all terms of the zero and the first orders in the Planck constant, Eqs. (\ref{eq7w})-(\ref{finalOmegw}) allow to determine an exact classical limit of quantum-mechanical equations of motion.

The new exact result (\ref{eq7w})-(\ref{finalOmegw}) significantly generalizes the earlier findings of Refs. \cite{gos,Quach1,Quach2}. The relativistic Hamiltonian derived in Ref. \cite{gos} by means of the Eriksen-Kolsrud transformation \cite{erik} (also called the exact FW transformation) does not provide for a simple transition to the classical limit except for the nonrelativistic approximation. This is caused by the difference between the FW and the Eriksen-Kolsrud representations \cite{PRD,JMPcond}. It should be also noted that the exact FW Hamiltonian found in Ref. \cite{gos} contains an error later reported in Ref. \cite{Quach1}. In Ref. \cite{Quach2}, a nonrelativistic Dirac particle interacting with a magnetic field in the presence of a gravitational wave has been considered and a possibility of the spin gravitational resonance has been mentioned. We analyze the results obtained in this work in Sec.~\ref{Weffects}.

The general equations (\ref{eq7w})-(\ref{finalOmegw}) clearly show that the gravitational wave causes the spin rotation even in the absence of the electromagnetic interactions and it can change the motion of the spin affected by these interactions. These properties are manifest in the Hamiltonian constituents ${\cal H}_{FW}^{(1)},{\cal H}_{FW}^{(2)}$ and ${\cal H}_{FW}^{(3)},{\cal H}_{FW}^{(4)}$, respectively. Our result confirms the conclusion made in Ref. \cite{gos} that the influence of the gravitational wave on the momentum and spin may be amplified by a sufficiently strong magnetic field. One can check that the significant improvement can take place when $\partial_c({\cal F}^b{}_a)/{\cal F}^e{}_d\ll \Omega_0$, where $\Omega_0$ is the frequency of the spin rotation caused by electromagnetic interactions. In some cases, this condition is equivalent to $\omega\ll \Omega_0$, where $\omega$ is the frequency of the gravitational wave. 

While exact gravitational wave solutions are known \cite{Brink1,Brink2,Brink3,rosen1937,einrosen,bondi0,bondi1,peres,pen1,pen2,griff,vdz,exact}, for most of practical purposes it is sufficient to consider the weak-field approximation. In this approximation, the spatial metric $\underline{g}{}_{AB} = h^{\widehat C}{}_Ah^{\widehat D}{}_B\delta_{{\widehat C}{\widehat D}}$ components are given by
\begin{equation}\label{nontrivial}
\underline{g}{}_{11}=1+2w_{\bigoplus},\qquad \underline{g}{}_{22}=1-2w_{\bigoplus},\qquad
\underline{g}{}_{12}=\underline{g}{}_{21}=2w_{\bigotimes}.
\end{equation}
Note an inaccuracy in the metric presented in Eq. (2) of Ref. \cite{Quach2}. The dynamics of the weak gravitational wave is described by the zweibein anholonomity object (\ref{PAB}) which in this approximation reads
\begin{equation}
\Phi^1{}_1 = - \,\Phi^2{}_2 = \partial_\sigma w_{\bigoplus},\qquad 
\Phi^1{}_2 = \Phi^2{}_1 = \partial_\sigma w_{\bigotimes}.\label{PABw}
\end{equation}  

For the electric and magnetic fields, we find $h = 1$, and (\ref{EEw})-(\ref{BBw}) reduce to
\begin{equation}\label{EEBBw}
\mathfrak{E}{}_a = (\delta_a^b - w^{b}{}_a)E_b,\qquad {\mathfrak{B}}{}^a = (\delta^a_b + w^{a}{}_b)B^b.
\end{equation} 

It is important to determine the gravitoelectromagnetic fields (\ref{ge}) and (\ref{gm}). For the Rosen-Bondi gravitational waves, they turn out to be transversal: $\mathcal{E}_{3} = \mathcal{B}^{3} = 0$ and 
\begin{equation}\label{gemw}
\mathcal{E}_{A} = -\,c\Phi_{AB}U^B,\qquad \mathcal{B}^{A} = -\,{\frac 1c}\,\eta^{AB}\mathcal{E}_{B}.
\end{equation}
In the weak-field approximation, the hats may be omitted, and $\Phi_{AB}$ is given by (\ref{EEBBw}). One should note that these fields are effective, they are defined not only by the gravitational field but also by the 4-velocity $U^\alpha$ of the particle. The quantum-mechanical counterparts of these equations are
\begin{equation}\label{expwvqm}
\mathcal{E}_{A} = -\,{\frac c{2m}}\left\{\Phi^B{}_{A}, \pi_B\right\},\qquad 
\mathcal{B}^{A} = -\,{\frac 1c}\,\eta^{AB}\mathcal{E}_{B},\qquad \mathcal{E}_{3} = \mathcal{B}^{3} = 0.
\end{equation}

The FW Hamiltonian of a charged fermion particle moving in the electromagnetic field and the weak gravitational wave can be presented in the general form
\begin{equation}
\begin{array}{c} {\cal H}_{FW}=\beta\epsilon' + e\Phi + {\frac {\hbar mc^2}{8}}\left\{
{\frac{1}{\epsilon'(\epsilon' + mc^2)}},\bm\Sigma\cdot(\bm\pi\times\bm{\mathcal{E}}
-\bm{\mathcal{E}}\times\bm\pi)\right\}\\ 
+ \,{\frac 14}\left\{\left(\frac{\mu_0mc^2}{\epsilon' + mc^2}+\mu'\right)\frac{1}{\epsilon'},
\Bigl[\bm\Sigma\cdot(\bm\pi\times\bm{\mathfrak{E}}-\bm{\mathfrak{E}}\times\bm\pi)-\hbar\nabla
\cdot\bm{\mathfrak{E}}\Bigr]\right\}\\  
- \,{\frac {\hbar mc^2}{4}}\left\{\frac{1}{\epsilon'}, \bm\Pi\cdot\bm{\mathcal{B}}\right\}
- {\frac 12}\left\{\left(\frac{\mu_0mc^2}{\epsilon'} + \mu'\right), \bm\Pi\cdot\bm{\mathfrak{B}}\right\}\\
+ \,\beta{\frac{\mu'}{4}}\left\{{\frac{c^2}{\epsilon'(\epsilon' + mc^2)}},
\Bigl[(\bm{\mathfrak{B}}\cdot\bm\pi)(\bm{\Sigma}\cdot\bm\pi)+ (\bm{\Sigma}
\cdot\bm\pi)(\bm\pi\cdot\bm{\mathfrak{B}}) + \frac{\hbar}{2}\left(\bm\pi\cdot\bm{\mathcal{J}}
+ \bm{\mathcal{J}}\cdot \bm\pi\right)\Bigr]\right\}.\end{array} \label{eq33new} 
\end{equation} 
Here $\mu_0 = {\frac {e\hbar}{2m}}$ is the Dirac magnetic moment and the gravitoelectric $\bm{\mathcal{E}}$ and the gravitomagnetic $\bm{\mathcal{B}}$ fields are given by Eqs. (\ref{expwvqm}). We can disregard the EDM while the AMM should be taken into account. 

Equation (\ref{eq33new}) underlies the investigation of the physical effects for the Dirac fermion in the weak gravitational wave.

\subsection{Analysis of physical effects in magnetic field}\label{Weffects}

Let us compare our result with the Hamiltonians obtained previously in \cite{gos,Quach1,Quach2}. In Ref. \cite{gos}, the Eriksen-Kolsrud transformation (also called the exact FW transformation) has been used. While this transformation is often exact, it is not convenient for deriving an unambiguous classical limit of the quantum-mechanical equations. Nevertheless, the Eriksen-Kolsrud and the FW transformations for the Dirac particle interacting with the magnetic field and moving in the gravitational wave lead to the same nonrelativistic Hamiltonians \cite{Quach1,Quach2}. To compare them with the exact Hamiltonian (\ref{eq33new}), we may take into account only the expansion up to the first and second orders in $v/c$ in terms proportional to the first and zeroth powers of $\hbar$, respectively. After omitting other terms, Eq.  (\ref{eq33new}) reduces to
\begin{eqnarray}
{\cal H}_{FW} &=& \beta mc^2+\left(\delta^{ab}-2w^{ab}\right)\frac{\pi_a\pi_b}{2m}\nonumber\\
&& - \,{\frac {\hbar}{2}}\,\bm\Pi\cdot\bm{\mathcal{B}} 
- \left(\mu_0 + \mu'\right)\bm\Pi\cdot\bm{\mathfrak{B}}.\label{eq30}
\end{eqnarray}

For understanding of the possible physical effects it is important to notice that the spin is affected by the anholonomic field $\bm{\mathfrak{B}}$ which bears an ``imprint'' (\ref{EEBBw}) of the gravitational wave on the applied magnetic field $\bm{B}$. As one knows, when a particle with a magnetic moment moves in the flat spacetime (no gravity) in a constant homogeneous magnetic field, its spin eigenstates defines the polarization along (or against) the applied field. However, when a rotating (or, in general, alternating) field is added in the plane perpendicular to the original constant field, the spin can flip and the highly interesting magnetic resonance phenomenon \cite{Rabi0,Rabi1,Schw,Bloch,Rabi2} occurs, which has numerous important applications. 

Supposing that the weak gravitational wave is an harmonic oscillatory process, one can qualitatively model this situation. To make the discussion more clear, let us assume that only one gravitational wave polarization is present, namely $w_{\bigoplus} = 0$, whereas 
\begin{equation}
w_{\bigotimes} = g_0\,\cos\varphi = g_0\,\cos\left(\omega t - \omega z/c\right),\label{wt}
\end{equation}
describing a wave with the frequency $\omega$ and amplitude $g_0$ propagating along the $z$-axis. 

Now, let us arrange the constant homogeneous magnetic field in the plane of the wave front: without loss of generality we can choose $\bm{B} = (B_0, 0, 0)$, where $B_0=\,$const. Then (\ref{EEBBw}) yields $\bm{\mathfrak{B}} = (B_0, B_0w_{\bigotimes}, 0)$, and we thus discover that the spin couples, see the last term in (\ref{eq30}), to the field configuration that reproduces the magnetic resonance conditions. Namely, the spin is affected by the constant homogeneous magnetic field along $x$ and an additional alternating field in the perpendicular plane $(y,z)$. The case when the alternating field is not rotating but performs a simple linear oscillation was first considered by Bloch and Siegert \cite{Bloch}, and using their results, we obtain the probability of the spin-flip  
\begin{equation}\label{Pf}
P_{-{\frac 12}} = {\frac {\sin^2\left\{\omega_0g_0(t - t_0)\Lambda/4\right\}}{\Lambda^2}}\,.
\end{equation}
This is the probability to find, at time $t$, the spin oriented oppositely to the initial state at $t_0$. Here 
\begin{equation}
\omega_0 = {\frac {2\left(\mu_0 + \mu'\right)B_0}{\hbar}}\label{Larmor}
\end{equation}
is the Larmor frequency, and 
\begin{eqnarray}
\Lambda^2 &=& 1 + {\frac {4(1 - \xi)^2}{g_0^2}},\label{lam}\\
\xi &=& {\frac {\omega}{\omega_0}}\left(1 - {\frac {g_0^2}{16\xi^2}}\right).\label{xi} 
\end{eqnarray}
This confirms the earlier conclusion \cite{Quach2} about the possibility to find manifestations of the gravitational wave with the help of the magnetic resonance type experiments.

One can notice that the effect described by Eqs. (\ref{Pf})-(\ref{xi}) is quadratic in the small quantity $g_0$, which is usual in the theory of magnetic resonance. At the same time, it would be of interest to search for the possible polarization effects which are linear in this quantity, and to analyze the spin components orthogonal to an initial spin polarization, along the lines of Ref. \cite{EPJC2017,epl}.

It is worthwhile to mention that in his computation Quach \cite{Quach2} rather paradoxically considered the case when the direction of the external magnetic field differs from the direction of the wave propagation by the small angle $\theta\rightarrow 0$, and this small parameter necessarily enters the final result. However, when $\theta = 0$ (the magnetic field along the wave propagation), we have $\bm{\mathfrak{B}} = \bm{B}$, and the gravitational effects disappear. In contrast, in our new derivation we consider a different setup when the magnetic field is applied in the wave front plane orthogonal to the wave propagation. Accordingly, no additional small parameters enter the results obtained, (\ref{Pf})-(\ref{xi}). 

\section{Discussion}\label{Conclusion}

In this paper we continue the study of the dynamics of the quantum and classical Dirac fermions with spin $1/2$ and dipole moments under the action of the gravitational and electromagnetic fields. The results obtained extend our earlier findings in \cite{PRD,PRD2,Warszawa,OST,OSTRONG,ostgrav} which were obtained for the case when the electromagnetic field was excluded whereas gravity was confined to the weak fields and the special static and stationary field configurations. We now considered the case of an arbitrary spacetime metric plus an arbitrary electromagnetic field. This is a nontrivial problem, because one cannot merely sum up the results describing the influence of the electromagnetic field on spin with the results for the gravity-spin effects. There is no direct superposition, since unlike the electromagnetism that couples only to electric charges and currents, gravity is universal and it couples to all kinds of matter, including the electromagnetic field. As a result, the action of gravity on spin is twofold: via the coframe and connection (\ref{AB1})-(\ref{AB3}) and via the electromagnetic field which gets modified Eqs. (\ref{EE})-(\ref{BB}) in the curved spacetime. 

In Sec.~\ref{Prelim}, we derived the Hermitian Dirac Hamiltonian (\ref{HamiltonDP}), and then in Sec.~\ref{Hamiltonian} we applied the Foldy-Wouthuysen transformation \cite{PRA} and constructed the Hamiltonian (\ref{eqFW})-(\ref{eq74}) in the FW representation for a particle  with spin $1/2$ and dipole moments in an arbitrary spacetime geometry and an arbitrary electromagnetic field. Making use of the FW Hamiltonian, we then derived the operator equations of motion, with a special focus on the quantum-mechanical equation of the spin motion (\ref{spinmeq}) and its semiclassical limit (\ref{dots}). These general results are important for the comparison of the dynamics of a quantum and classical spinning particle in external fields, when addressing the issue of the validity of the equivalence principle. In Sec.~\ref{CS}, we considered the general theory of the classical particle with spin in external fields and applied it to the analysis of spin dynamics under the joint action of gravity and electromagnetism in Sec. \ref{CSQ}, where we derived the expressions (\ref{Omtot}) and (\ref{OmegaT}) for the angular velocity of spin precession in the general inertial, gravitational and electromagnetic fields. Our main result is the demonstration that the classical spin dynamics is completely consistent with the semiclassical quantum dynamics of the Dirac fermion in an arbitrary curved spacetime and any electromagnetic field. 

Finally, in Sec. \ref{SG} we used our general formalism to derive the theoretical framework for the investigation of the dynamics of quantum spinning particle in the field of a gravitational wave. In particular, we are able to confirm and improve the earlier findings \cite{Quach2} on the theoretical possibility of using the magnetic resonance type setup to reveal gravitational wave effects on a spin.

\section*{Acknowledgments}

We thank F. W. Hehl for reading the preliminary
draft and for his valuable comments and discussions. The work was supported in part by the
Belarusian Republican Foundation for Fundamental Research (Grant No. $\Phi$16D-004),
by the Heisenberg-Landau program of the German Ministry for Science and Technology
(Bundesministerium f\"{u}r Bildung und Forschung), by PIER (``Partnership for Innovation,
Education and Research'' between DESY and Universit\"at Hamburg), and
by the Russian Foundation for Basic Research (Grants No. 14-01-00647 and No. 16-02-00844-A).

\end{document}